\documentclass[twocolumn]{autart}

\RequirePackage{amsmath, amssymb, graphicx, url,mathtools,
xspace,subfigure,braket,epstopdf}
\usepackage{comment} 
\usepackage[makeroom]{cancel}

\usepackage{graphicx,subfigure}
\usepackage{epsfig} 
\usepackage{mathptmx} 
\usepackage{times} 
\usepackage{amsmath} 
\usepackage{amssymb}  
\usepackage{amsfonts}
\usepackage{euscript}
\usepackage{color}

\begin{document}

\begin{frontmatter}

\thanks[footnoteinfo]{
This paper was
not presented at any IFAC meeting.
Corresponding author Gianluigi
Pillonetto Ph. +390498277607.}

\title{Estimation of sparse linear dynamic networks\\ using the stable spline horseshoe prior}

\author[First]{Gianluigi Pillonetto}
\address[First]{Department of Information  Engineering, University of Padova, Padova, Italy (e-mail: giapi@dei.unipd.it)}


\maketitle
\begin{abstract}
Identification of the so-called dynamic networks is one of the most challenging problems appeared recently in control literature.
Such systems consist of large-scale interconnected systems, also called modules. 
To recover full networks dynamics the two crucial steps are \emph{topology detection}, where one has to infer from data 
which connections are active, 
and \emph{modules estimation}. 
Since a small
percentage of connections are effective
in many real systems, the problem finds also fundamental connections
with group-sparse estimation. In particular, in the linear setting 
modules correspond to unknown impulse responses 
expected to have null norm but in a small fraction of samples. 
This paper introduces a new Bayesian approach for 
linear dynamic networks identification where impulse responses 
are described through the combination of two particular prior distributions. 
The first one is a block version of the horseshoe prior, a model
possessing important global-local shrinkage features. The second one
is the stable spline prior, that encodes information on smooth-exponential decay of the modules.
The resulting model is called stable spline horseshoe (SSH) prior.
It implements aggressive shrinkage of small impulse responses 
while larger impulse responses are conveniently subject to stable spline
regularization. Inference is performed by a Markov Chain Monte Carlo scheme,
tailored to the dynamic context and able to efficiently return the posterior of the modules in sampled form. 
We include numerical studies that show how
the new approach can accurately reconstruct sparse networks dynamics 
also when thousands of unknown impulse response coefficients
must be inferred from data sets of relatively small size.
\end{abstract}

\end{frontmatter}

\section{Introduction}

Modeling complex physical systems is crucial in 
several fields of science and engineering, including also biomedicine and 
neuroscience \cite{Hagmann2008,Hickman2017,Pagani2013,Prando2020}.
Such systems can often be seen as \emph{dynamic networks}, i.e.
as a large set of interconnected dynamic systems.
They can be also seen as a set of nodes associated to measurable (noisy) outputs,
able to communicate with other nodes through modules driven by 
measurable inputs or noises. Identification of such systems poses several interesting challenges.
First, they are often large-scale and network topology is typically unknown  \cite{Chiuso2012,Materassi2010,VdH2013}.
Hence, one has often to postulate the existence of many connections 
and then to understand from data which are really working. The estimation process
has also to be informed that real physical systems often 
contain a limited number of active links. 
Only the significant ones should be detected and this
fact connects the problem with sparse estimation \cite{Hastie01}.
Beyond the knowledge of network topology, 
accurate estimation of modules dynamics is then needed and
also this step is far from trivial.
The use of parametric models to describe the modules, like 
rational transfer functions, can 
lead to high-dimensional non convex optimization problems \cite{Ljung:99,Soderstrom}. 
They can also require model (discrete) order selection 
whose computational cost is combinatorial.\\

In the context of \emph{linear dynamic networks},
where modules are defined by impulse responses,  
many numerical procedures have been recently proposed to overcome 
the identification issues mentioned above.
 Approaches based on local  multi-input-single-output (MISO) models 
 can be found in \cite{Dankers2016,Everitt2018,Materassi2020}. 
Contributions relying on variational Bayesian inference and/or nonparametric regularization,
connected with the new technique discussed later on in this paper,  
are described in \cite{Jin2020,Ramaswamy2021,Yue2021}.
Methods to infer the full network dynamics using (structured) multi-input-multi-output (MIMO) models 
are instead reported in \cite{Weerts2018,Fonken2020}, 
with estimates consistency 
studied in \cite{Ramaswamy2021b} (also assuming noise correlation over different nodes). 
See  also \cite{Bazanella2017,Goncalves2008,Hendrickx2019,Weerts2018b}
for insights on identifiability issues and \cite{Hayden2016}
where compressed sensing is exploited.\\

In this paper we introduce a new approach to estimation
of linear and stable dynamic networks within the framework of Bayesian regularization.
Our estimator includes the following information on the network:
block-sparsity,
i.e. many impulse responses are expected to have null norm,
and smooth exponential decay, i.e. any module is assumed to be 
an exponentially stable dynamic system.
Our model is obtained by combining two distributions that have recently played
important but distinct roles in sparse estimation and system identification: 
the \emph{horseshoe} and the \emph{stable spline} prior.\\

To briefly overview the \emph{horseshoe prior}, 
first it is useful to recall that  Lasso is one of the most important approaches 
introduced for sparse estimation \cite{Lasso1996,LARS2004}.
It recovers the scalar coefficients of an unknown signal as the solution
of a regularized optimization problem, using the $\ell_1$ norm as penalty.
This approach enjoys a stochastic interpretation, the so called Bayesian Lasso \cite{Park2008}, 
where the unknown parameters are independent Laplacian random variables.
It is equivalent, and more convenient for our discussion, to 
use a representation in terms of a scale mixture of normals. 
For known variance $\lambda^2_i$, the Bayesian Lasso sees the parameters 
as uncorrelated Gaussians $\mathcal{N}(0,\lambda^2_i)$, with 
the $\lambda^2_i$ mutually independent and following a common exponential distribution\footnote{This kind of hierarchy describes also the Bayesian interpretation of the relevance vector machine \cite{Tipping2001} just using inverse-gamma (in place of exponential) mixing to describe the variances.}. 
This representation allows an efficient implementation by a Markov chain Monte Carlo scheme \cite{Gilks}. 
However, a limitation of this approach
(and of all the Lasso-type regularizers) is that uniform shrinkage is provided to all the coefficients.
Hence, it can be difficult to promote a high level of sparsity 
while simultaneously avoiding oversmoothing of large coefficients.
The horseshoe prior introduced in \cite{Carvalho2009,Carvalho2010}
circumvents this problem by using
$\mathcal{N}(0,\tau^2 \lambda^2_i)$ as conditional density.
Furthermore, half-Cauchy distributions are assigned to all the 
standard deviations $\lambda_i$ and $\tau$ that 
represent, respectively, the local and the global 
shrinkage parameters. In particular, $\tau$ regulates the global sparsity 
level so that  its estimate will be very small in nearly black objects.
Then, the half-Cauchy heavy tails allow the local $\lambda_i$
to assume values large enough to retain the  
relevant signal components, letting them be virtually unaffected by the aggressive shrinkage induced 
by $\tau$ (see also Fig. 2 in \cite{Carvalho2009} for an illuminating description 
of the different sparsity features of Bayesian Lasso and horseshoe). 
In this way, the horseshoe estimator not only has mean squared error properties of global shrinkage estimators but 
is also asymptotically minimax for recovering nearly black objects  \cite{Donoho1992,Polson2012,Pas2014}.
See also \cite{Johndrow2020} and references therein for discussions on implementation issues and 
a new Markov Chain Monte Carlo (MCMC) scheme for inference in high-dimensions.\\

The \emph{stable spline prior} was instead proposed in \cite{SS2010}
and provides a nonparametric description 
of stable linear systems. The idea there developed was to search for the unknown impulse response in a high-dimensional space given by a stable reproducing kernel Hilbert space \cite{Aronszajn50,AbsSum2020,MathFoundStable2020,Scholkopf01b}, 
 with ill-posedness circumvented through penalty terms tailored for dynamic systems.
Also this approach admits a Bayesian interpretation where the impulse response 
is a zero-mean Gaussian vector with covariance 
given by the stable spline kernel. This latter 
includes information on smooth exponential decay.
In particular, the 
impulse response variance decay rate is regulated by  
the kernel parameter $\alpha$.  
A maximum entropy derivation of this prior can be found in \cite{CACCLP16}, while
other important classes of covariances to model dynamic systems are e.g. described in \cite{ChenOL12,CALCP14,Pillonetto2016,PillonettoHybrid,Chen19,Darwish2018}.\\  
The use of the stable spline prior for linear system identification has shown some important advantages in comparison
with other classical techniques like parametric prediction error methods \cite{Ljung:99}. It has been shown that replacing classical model discrete order selection with continuous kernel parameter tuning 
may better trade-off bias and variance. Hence, impulse response estimators with more favourable mean squared error properties can be achieved \cite{SurveyKBsysid}. Minimax properties of the stable spline estimator have been also
recently derived in \cite{PillonettoTAC2021}, outlining their dependence on the parameter $\alpha$.\\

The new Bayesian approach for identification of linear dynamic networks
introduced in this paper relies on a \emph{combination of a block version of the horseshoe prior and the stable spline kernel}.
To jointly perform topology detection and modules estimation, we model
the impulse responses as independent zero-mean Gaussian vectors.
Their covariances depend on local parameters $\lambda_k$, where
$k$ indexes the modules, and on
two global parameters $\tau,\alpha$. 
The scalar $\tau$ now regulates the level of block-sparsity 
of the network, hence its estimate will be very small if many modules are null.
The product $\tau^2 \lambda^2_k$ then defines the scale factor of the 
stable spline kernel associated to the $k$-th module.
Finally, the parameter $\alpha$ is related to the maximum absolute value of 
 the dominant poles of the impulse responses present in the network.
This new model, that we call
the \emph{stable spline horseshoe (SSH)} prior, 
thus provides aggressive shrinkage of small impulse responses 
while larger impulse responses are conveniently subject to stable spline
regularization.\\
The hierarchy underlying our model, 
combined with the representation of Cauchy distributions as a mixture of inverse-gamma  distributions
described in \cite{Makalic2016},
permits 
the use of a particular MCMC inference scheme, 
known as Gibbs sampling \cite{Gilks}.
Our procedure is especially efficient since it
is tailored to the system identification context where
the size of each impulse response is typically much smaller 
than the number of measurable outputs. 
For known $\alpha$, 
it is sufficient to compute off-line 
inverses and Cholesky factors of small matrices 
to generate samples 
from the posterior of the modules with low computational cost. 
The concept of Bayesian evidence can then be exploited 
to learn also $\alpha$ from data if  no enough
information is available to fix its value a priori.
Numerical studies will illustrate that our procedure 
permits an accurate reconstruction of sparse networks  
also when thousands of unknowns 
must be inferred from data sets of relatively small size.\\

The paper is so organized.
Section \ref{Section2} reports the problem statement.
Section \ref{Section3} describes SSH.
The MCMC scheme to perform sparse estimation of dynamic networks 
using SSH is described in Section \ref{Section4}.
Numerical experiments are reported in Section \ref{Section5}.
Conclusions then end the paper.

\section{Problem statement}\label{Section2}

Given two discrete-time signals 
$f$ and $g$, we denote with $(f \otimes g)_i$
their convolution evaluated at the sampling instant $t_i$.
Then, each node in the network is associated with the following
measurements model
\begin{equation}\label{MM1}
y_i = \left( \sum_{k=1}^p (f_k \otimes u_k)_i \right) + e_i, \quad i=1,\ldots,n
\end{equation}
where $y_i$ is the output measured at $t_i$, the causal signal $f_k$ is the 
impulse response connected with the $k$-th module,
the signal $u_k$ is the measurable input entering the $k$-th module.
Finally, the noises $e_i$ are zero-mean independent Gaussians
of variance $\sigma^2$.\\  
Our problem is to estimate the $p$ impulse responses $\{f_k\}_{k=1}^p$
using the data $\{y_i\}_{i=1}^n$ and the known set of inputs $\{u_k\}_{k=1}^p$.
The number of modules $p$ may be large but it is known that only
a small fraction of the impulse responses has norm different from zero. A
group-sparse estimation problem thus arises.\\
It is useful to rewrite 
\eqref{MM1} in matrix-vector form by assuming that 
FIR models of length $m$ can capture the dynamics of each module.
After building suitable Toeplitz matrices $G_k \in \mathbb{R}^{n \times m}$ with
inputs values, the measurements model becomes
\begin{equation}\label{MM2}
Y = \left( \sum_{k=1}^p G_k \theta_k \right) + E
\end{equation}
where the column vectors $Y,\theta_k$ and $E$ contain, respectively,
the $n$ output measurements, the $m$ impulse impulse response coefficients defining the 
$k$-th module and the $n$ Gaussian noises.\\
It has been implicitly assumed that 
all the signals $u_k$ defining the $G_k$
contain only exogenous variables. Actually, feedback could be also present
so that some inputs could embed also endogenous variables.
In this case, \eqref{MM2} can form a VARX model
where some regression matrices are defined also through past outputs.
This does not influence any of our algorithmic developments.
Simply, the estimates of $\theta_k$ become estimates of the 
one-step ahead predictors associated to the modules, from which
predictors over any desired horizon can be easily obtained \cite{Ljung:99}.

\section{The stable spline horseshoe prior}\label{Section3}

\subsection{The horseshoe prior}

For the moment, let us just consider the vector $\theta_k$
containing the FIR coefficients  associated to the $k$-th module,
with $\theta_{ki}$ to denote its $i$-th element.
Adopting the horseshoe prior, 
conditional on their variances,
all of these components 
are seen as independent zero-mean Gaussians, i.e.
\begin{equation}\label{HSprior}
\theta_{ki} \Big |  \tau^2,\lambda^2_{ki} \ \sim \ \mathcal{N}\big(0,\tau^2 \lambda^2_{ki}\big)
\end{equation}
where $\tau^2$ and $\lambda^2_{ki}$ are
the global and local shrinkage parameters already 
mentioned in Introduction. The prior model for 
$\theta_k$ is then completed by assuming
that such variances are independent random variables with distributions
defined by
\begin{equation}\label{HSpriorVar}
\lambda_{ki} \ \sim \ \mathcal{C}^+(0,1), \quad \tau \ \sim \ \mathcal{C}^+(0,1).
\end{equation}
Here, $\mathcal{C}^+(0,1)$ denotes the standard half-Cauchy distribution
whose probability density function is given by
$$
\mathbf{p}(x) = \frac{2}{\pi(1+x^2)}, \quad x>0.
$$

\subsection{The stable spline horseshoe prior}

Even if the horseshoe prior  (\ref{HSprior},\ref{HSpriorVar})
possesses important sparsity features, 
it is in general a poor statistical description of a stable linear system.
In fact, it models the impulse response coefficients
as stationary white noise, not embedding any information
on smooth exponential decay. Instead, good mean squared error properties
in system identification require the introduction of suitable correlation 
among the samples, as also described in a deterministic setting in \cite{ChenOL12}.
Furthermore, the global shrinkage parameter $\tau$ entering 
(\ref{HSprior},\ref{HSpriorVar}) describes
the level of sparsity of impulse response coefficients.
But, in dynamic networks, one would rather use it to control group-sparsity, 
i.e. the number of modules (network connections) that are null.\\
To face the above problems, we now introduce the stable spline horseshoe prior.
Conditional on the knowledge of its $m \times m$ covariance $\Sigma_k$, the impulse response $\theta_k$ 
is still seen as a zero-mean Gaussian vector, hence one has
\begin{equation}\label{SSHSprior}
\theta_k \Big |  \Sigma_k \ \sim \ \mathcal{N}\big(0,\Sigma_k \big).
\end{equation}
The main novelty in this work is now the definition of the random matrix $\Sigma_k$
through the stable spline kernel and a scale factor related to the horseshoe prior.
In particular, let $K$ be the $m \times m$ matrix  whose $i,j$ entry is
\begin{equation}\label{SS}
K_{ij} = \alpha^{\max(i-1,j-1)}, \quad 0\leq \alpha<1.
\end{equation}
This is exactly the covariance induced by the first-order stable spline model 
where the kernel parameter $\alpha$ regulates how fast the  impulse
response variance is expected to decay to zero.
Then, the covariance of $\theta_k$ is given by
\begin{equation}\label{SSHSpriorVar1}
\Sigma_k =  \tau^2 \lambda^2_k K 
\end{equation}
where the scale factor $\tau^2 \lambda^2_k$ is a random
variable with statistics defined by
\begin{equation}\label{HSpriorVar2}
\lambda_k \ \sim \ \mathcal{C}^+(0,1), \quad \tau \ \sim \ \mathcal{C}^+(0,1).
\end{equation}
Our SSH model is so completely defined by  (\ref{SSHSprior}-\ref{HSpriorVar2}).
Note that, differently from (\ref{HSprior},\ref{HSpriorVar}) where
the distribution of each $i$-th impulse response coefficient of the $k$-th module 
was function of a specific $\lambda_{ki}$,
any $\theta_k$ is now associated with only one local shrinkage parameter $\lambda_k$.
So, products $\tau^2\lambda^2_k$ close to zero 
will force the entire $k$-th module to have norm close to zero.
This promotes block-sparsity whose level is controlled by the global shrinkage parameter  $\tau$
that thus now establishes how many connections are not significant. 
Next, when a module is detected to be relevant,
the stable spline kernel $K$ will regularize its estimate.
In particular, information
on smooth-exponential decay is given through $\alpha$
whose value  
is connected with the maximum absolute value of 
the dominant poles of the modules.
Variations of such model are also possible, e.g. a more complex description 
can associate to each $\theta_k$ a specific kernel parameter 
$\alpha_k$.\\

\subsection{Graphical illustrations of SSH}

\begin{figure}
\center {\includegraphics[scale=0.4]{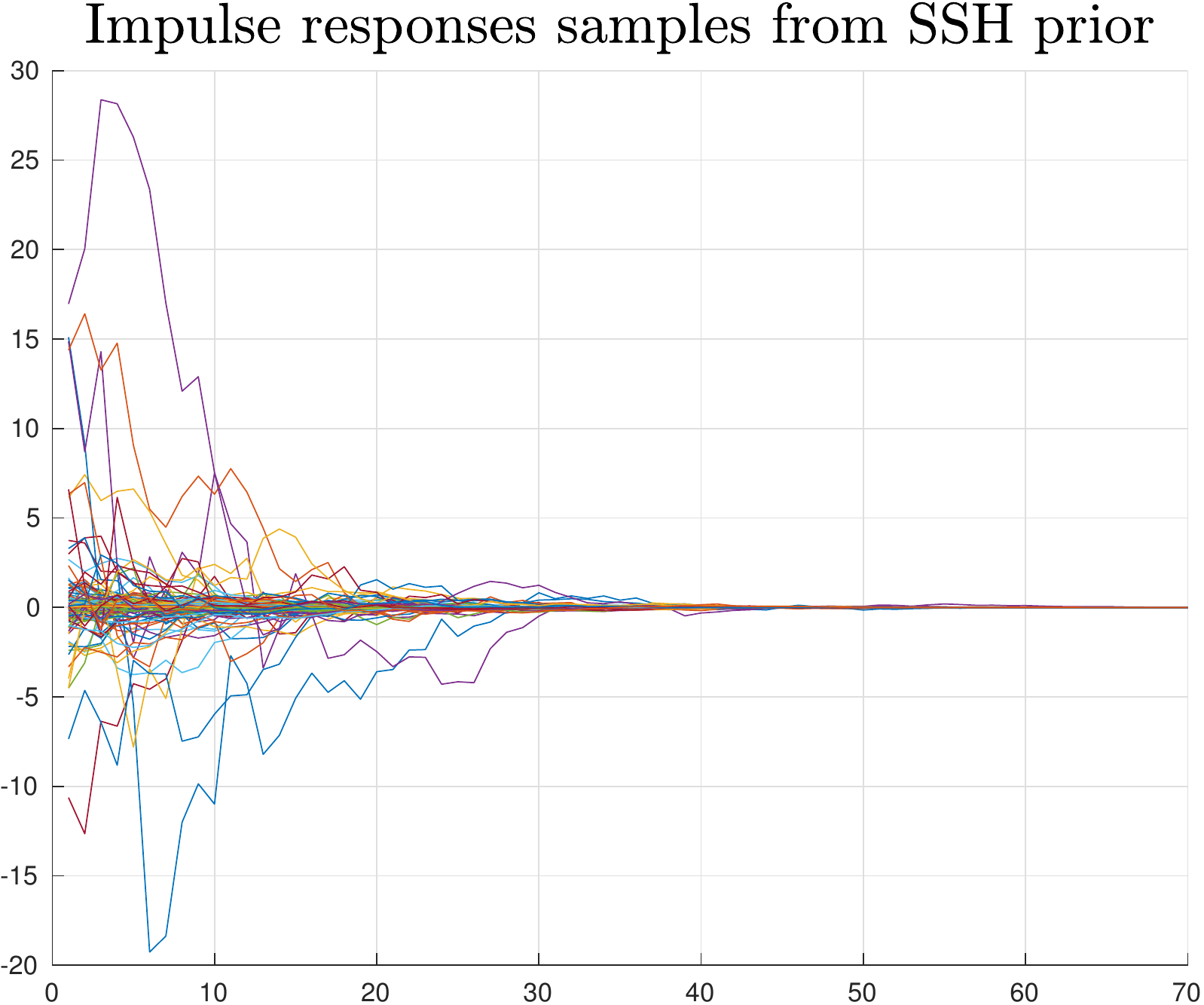}}
\caption{Samples from the stable spline horseshoe prior with $\tau=1$ and $\alpha=0.8$.}
\label{Fig1}
\end{figure}

\begin{figure*}
{\includegraphics[scale=0.35]{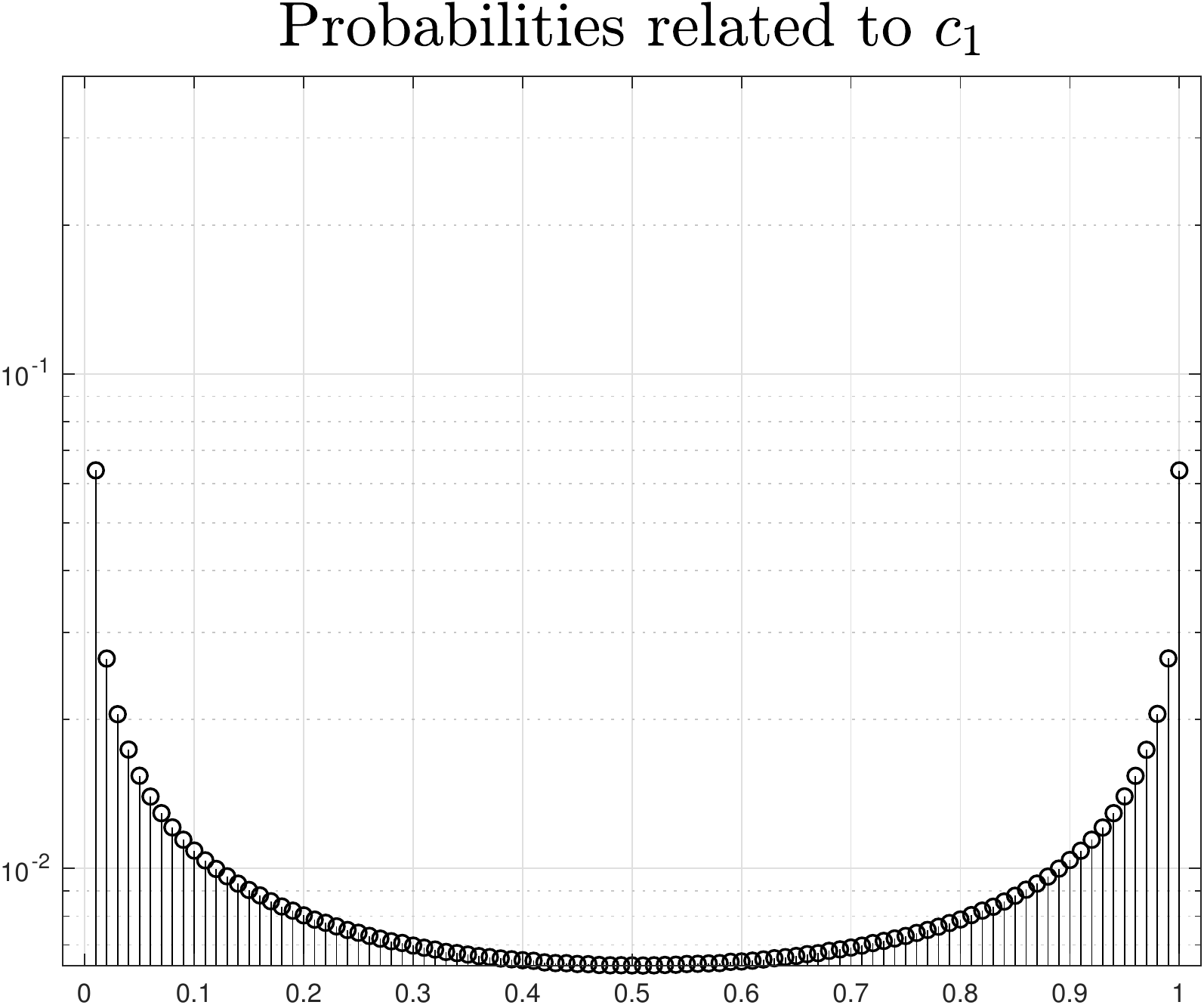}} \ {\includegraphics[scale=0.35]{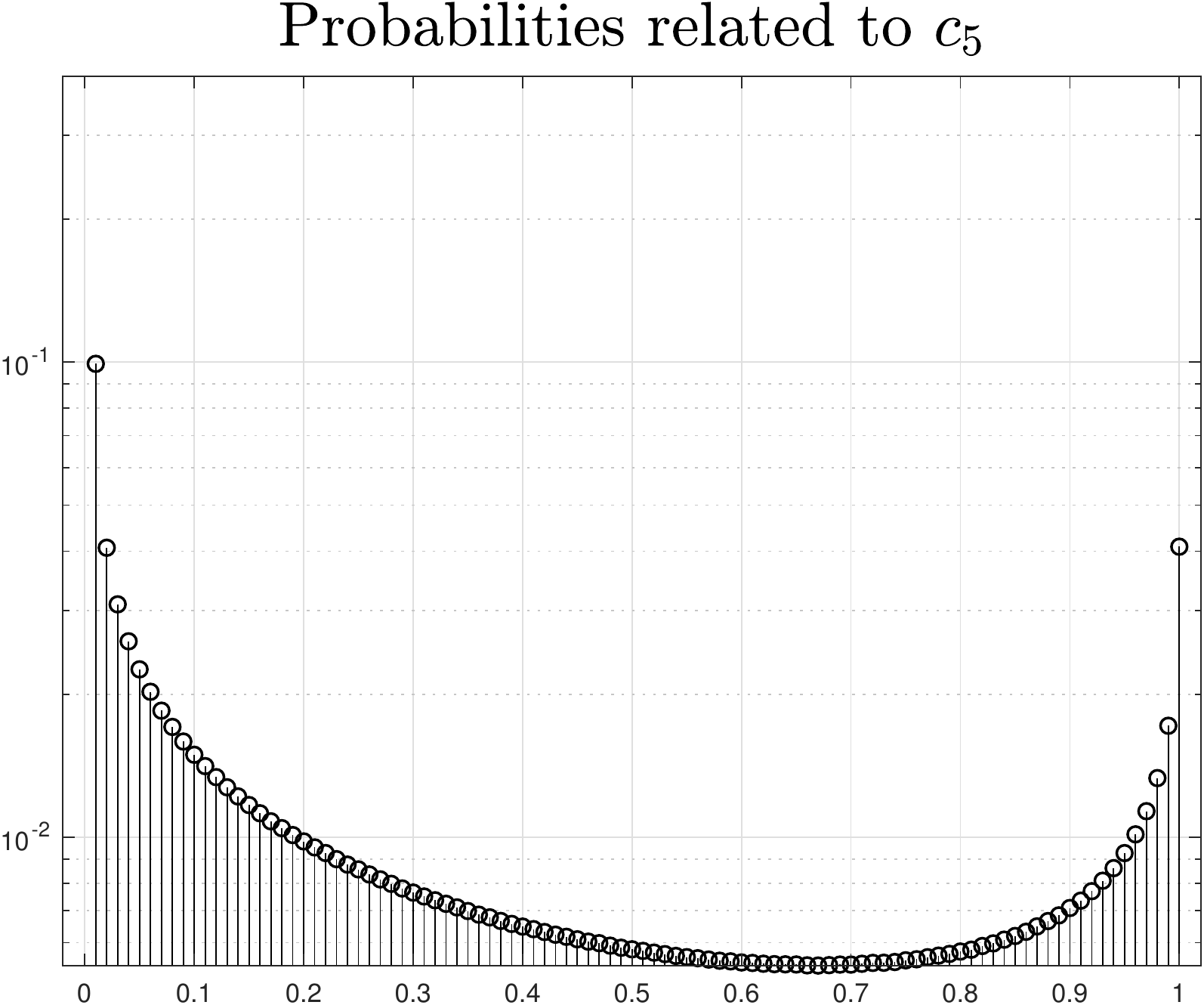}} \ {\includegraphics[scale=0.35]{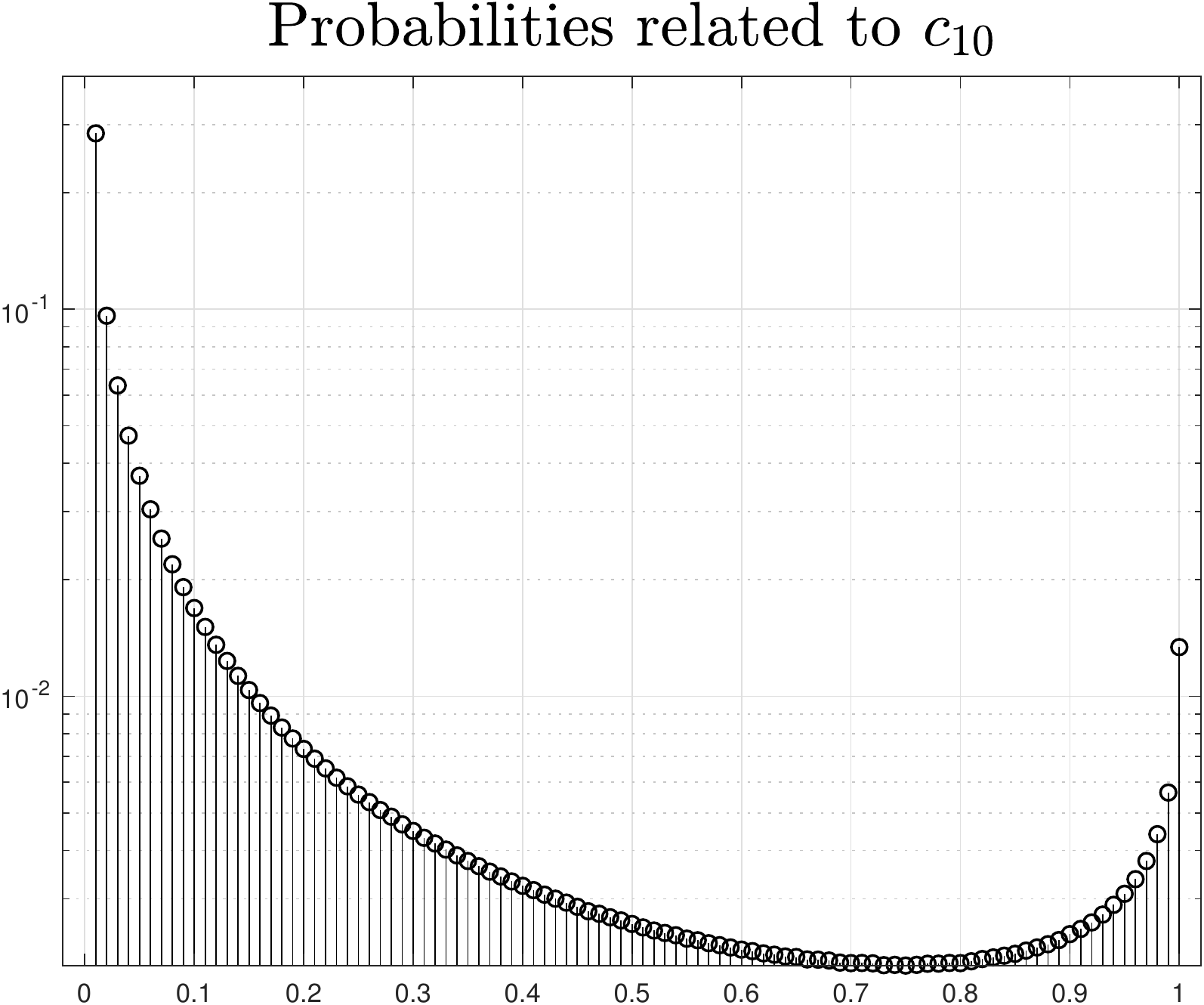}}
\caption{Probabilities associated to the shrinkage parameters $c_1,c_5$ and $c_{10}$ as defined in 
\eqref{ShrCoef} with 
$\alpha=0.8$.
For each value $x=\{0.01,0.02,\ldots,1\}$ the figure reports on the $y$-axis the probability of the event
$x-0.01 < c_i < x$.}
\label{Fig2}
\end{figure*}

We provide some illustrations of the prior 
defined in (\ref{SSHSprior}-\ref{HSpriorVar2}) by setting $\tau=1$
and $\alpha=0.8$.\\ 
We draw 100 candidate impulse responses 
from SSH. Each of them is obtained by generating 
a realization of $\lambda_k$ from the standard half-Cauchy distribution.
This then defines the covariance \eqref{SSHSpriorVar1}
from which a sample of $\theta_k$ from a multivariate Gaussian 
with stable spline covariance is drawn. The realizations are plotted in Fig. \ref{Fig1}
and two key features can be observed. First, all of them 
are quite smooth, with an exponential decay to zero
induced by the parameter $\alpha$. Second, there is a form of dichotomy
underlying the realizations. Some of them have norm close to zero, while
other ones can assume quite large values thanks to the flat tails of the 
half-Cauchy distribution. Indeed, the variance of $\lambda_k$ is infinite and this
can allow some impulse responses to be relevant even in presence of a small
$\tau$ value.\\
The second illustration considers a particular system identification problem
where there is a single module $\theta_k$ and the input is a unit impulse. 
Hence, direct noisy samples
of the impulse response can be observed, i.e.
$$
y_i = \theta_{ki} + e_i,
$$
with the noises $e_i$ of unit variance.
Under the SSH prior, it is easy to see that
the minimum variance estimator of each coefficient $\theta_{ki}$ conditional
on $\lambda_k$ and the sole output $y_i$ is
\begin{equation}\label{ShrCoef}
\mathbb{E}\big(\theta_{ki} \big | y_i,\lambda_i \big)  = c_i y_i, \quad c_i  \coloneqq \frac{\alpha^{i-1} \lambda_k^2}{1+\alpha^{i-1} \lambda_k^2}.
\end{equation}
The $c_i$ is thus a stochastic shrinkage coefficient being 
function of the random variable $\lambda_k$. 
It assumes values on the unit interval: 
$c_i=0$ and $c_i=1$ provide, respectively, a null or unbiased estimate.
Even if, conditional only on $y_i$, one then has
$$
\mathbb{E}\big(\theta_{ki} \big | y_i \big)  = \mathbb{E} \big(c_i \big | y_i \big) y_i,
$$
it is quite useful to plot the distribution of $c_i$ before seeing any data 
to understand which kind of impulse responses 
the SSH estimator expects to see. 
For $i=1$ such distribution is especially simple 
(as described a few lines below) while,
in general, it can be reconstructed in sampled form for any $i$.
In fact, after setting $\alpha=0.8$, one can draw
realizations of $c_i$ by drawing samples of
$\lambda_k$ from the half-Cauchy and then computing the ratios 
$\alpha^{i-1} \lambda_k^2/(1+\alpha^{i-1} \lambda_k^2)$.\\
When $i=1$ the kernel parameter $\alpha$ has no effect and
the shrinkage weight $c_1$ coincides with that induced by the horseshoe prior.  
It thus follows a beta probability density function with parameters 0.5, symmetric over the unit interval \cite{Carvalho2009}[Section 2.1]. 
Since such pdf is unbounded at the boundaries
of the unit interval, to ease the comparison with the other cases it is convenient to display the probability
of the events $c_1 \in (x-\Delta,x)$. This is done in the left panel of Fig. \ref{Fig2} 
setting $\Delta=0.01$ and $x=\{0.01,0.02,\ldots,1\}$.
The shrinkage profile is symmetric, hence the estimator
expects a priori to see either a strong first impulse response coefficient
or a null one.\\
As the time $i$ progresses, the dynamics underlying
the stable spline prior come into play. 
The shape changes: $\alpha$ assigns 
increasingly probability
to null values of the coefficients in the impulse response tail.
This phenomenon is illustrated in the middle and right panels of
Fig.  \ref{Fig2} that display probabilities associated to $c_5$ and $c_{10}$
adopting the same rationale followed to build the left panel.
The effect of stability information
on the horseshoe prior is evident.

\section{Full Bayesian model and MCMC identification}\label{Section4}

\subsection{Bayesian network model}

For the moment, the kernel parameter $\alpha$ is assumed known and
the dependence of some variables on its value is omitted to simplify notation.\\ 
The random variables entering our Bayesian model are the data $Y$,
the impulse responses $\{\theta_k\}_{k=1}^p$, the mutually independent variances 
$\{\lambda_k\}_{k=1}^p,\tau$ and $\sigma^2$. 
Following \cite{Makalic2016}, we also introduce
the auxiliary independent random variables $\{\nu_k\}_{k=1}^p$ and $\xi$ to  
represent the Cauchy distributions of the local and global shrinkage parameters
as a mixture of inverse-gamma distributions.
In particular, $\mathcal{I}_g(a,b)$ 
denotes the inverse-gamma with probability density function
$$
\mathbf{p}(x) = \frac{b^a}{\Gamma(a)}x^{-a-1}e^{-b/x}, \quad x>0.
$$ 
Under the SSH prior, our Bayesian model is then fully defined by the following
conditional distributions
$$
Y \big | \big( \{\theta_k\}_{k=1}^p,\sigma^2,\{\lambda_k,\nu_k\}_{k=1}^p,\tau,\xi \big) \sim \mathcal{N}\Big( \sum_{k=1}^p G_k \theta_k , \sigma^2 I_n\Big),
$$
$$
\{\theta_k\}_{k=1}^p \big | \big( \sigma^2,\{\lambda_k,\nu_k\}_{k=1}^p,\tau,\xi \big)  \sim \prod_{k=1}^p \mathcal{N}\Big( 0, \tau^2 \lambda^2_k K \Big),
$$
$$
\lambda^2_k \big | \nu_k  \sim \mathcal{I}_g(1/2,1/\nu_k),
$$
$$
\tau^2 \big |  \xi  \sim \mathcal{I}_g(1/2,1/\xi),
$$
where $\{\lambda^2_k \big | \nu_k\}_{k=1}^p$ and $\tau^2 \big |  \xi$ are all mutually independent,  
$$
\nu_1,\ldots,\nu_p,\xi  \sim \mathcal{I}_g(1/2,1),
$$
$$
\sigma^2 \sim \frac{d\sigma^2}{\sigma^2} \quad \text{(Jeffrey's prior)}.
$$

\subsection{Gibbs sampling for dynamic networks estimation}

It is useful to indicate with $\mathcal{V}$ the random vector containing all the unknown variables
given by $\{\theta_k\}_{k=1}^p$, $\sigma^2$,
$\{\lambda_k\}_{k=1}^p,\tau,\{\nu_k\}_{k=1}^p$ and $\xi$.
Below, it is then shown that the probability density function of $\mathcal{V}$ conditional on $Y$
can be reconstructed in sampled form in an efficient way through Gibbs sampling.
This particular MCMC scheme requires to draw samples 
sequentially from the simple full-conditional
distributions reported below.\\
For what regards the modules, we update separately each impulse response $\theta_k$.
One has
$$
\theta_k \big | \big( \mathcal{V} \setminus  \theta_k, Y  \big) \sim  \mathcal{N}(\hat{\mu}_k,\hat{\Sigma}_k)
$$
where
$$
\hat{\Sigma}_k = \left( \frac{G^\top_kG_k}{\sigma^2}+ \frac{K^{-1}}{\tau^2 \lambda^2_k}   \right)^{-1}
$$
$$
\hat{\mu}_k = \hat{\Sigma}_k \frac{G^\top_k}{\sigma^2}  \left(Y-  \sum_{j \neq k} G_j \theta_j \right)
$$
The variances $\sigma^2,\lambda_k^2$ and $\tau^2$ change at any MCMC iteration.
Hence, drawing samples of $\theta_k$ requires computation of new inverses
and decompositions of $\hat{\Sigma}_k$. 
However, we can rewrite the posterior covariance as follows
$$
\hat{\Sigma}_k = \sigma^2 K^{1/2}\left( K^{1/2} G^\top_kG_k K^{1/2} +  \frac{\sigma^2}{\tau^2 \lambda^2_k} I_m   \right)^{-1} K^{1/2},
$$
and, for any module $k$, compute the following SVD just once (before running the chain):
$$
K^{1/2} G^\top_kG_k K^{1/2} = U_k D_k U_k^\top.
$$
Then, we obtain
$$
\hat{\Sigma}_k = A_kA_k^\top 
$$
with
$$
A_k= \sigma K^{1/2} U_k \left(D_k + \frac{\sigma^2}{\tau^2 \lambda^2_k} I_m\right)^{-0.5}.
$$
This provides a closed-form expression of 
the posterior mean $\hat{\mu}_k $ and covariance $\hat{\Sigma}_k$  as a function of the 
$\sigma^2$ and the shrinking parameters. 
Samples from the conditional density of $\theta_k$ are then efficiently obtained as
$$
\hat{\mu}_k + A_k \omega_k
$$
where $\omega_k$ contains independent zero-mean Gaussians of unit variance.\\ 

We will see that all the remaining unknown variables 
can be updated just sampling from inverse-gamma distributions.\\ 
For what regards the noise variance, it holds that
$$
\sigma^2 \big | \big( \mathcal{V} \setminus  \sigma^2, Y \big)  \sim  \mathcal{I}_g \Big(\frac{n}{2}, \frac{1}{2} \big\| Y - \sum_{k=1}^p G_k \theta_k \big\|^2  \Big).
$$
We now consider the conditional distribution of the local and global shrinkage parameters 
given, respectively, by $\lambda^2_k$ and $\tau^2$.
Given a column vector $v$, let
$$
\| v \|^2_{K} = v^\top K^{-1} v
$$
where $K$ is the stable spline matrix defined in \eqref{SS}.
Then, after  some calculations one obtains 
$$
\lambda^2_k \big | \big( \mathcal{V} \setminus  \lambda_k, Y \big)  \sim   \mathcal{I}_g \Big(\frac{m+1}{2},  
\frac{1}{\nu_k} + \frac{\| \theta_k \|^2_{K} }{2\tau^2} \Big)
 $$
and
 $$
\tau^2 \big | \big( \mathcal{V} \setminus  \tau, Y \big)  \sim   \mathcal{I}_g \Big(\frac{mp+1}{2}, \frac{1}{\xi}  + 
\sum_{k=1}^p \frac{\| \theta_k \|^2_{K} }{2\lambda_k^2}\Big).
 $$

Finally, the conditional distributions of the auxiliary variables remain identical to those reported
in \cite{Makalic2016}, i.e.
$$
\nu_k \big | \big( \mathcal{V} \setminus  \nu_k, Y \big)  \sim  
\mathcal{I}_g \Big(1, 1 + \frac{1}{\lambda_k^2} \Big).
$$
and
$$
\xi \big | \big( \mathcal{V} \setminus  \xi, Y  \big) \sim  
\mathcal{I}_g \Big(1, 1 + \frac{1}{\tau^2} \Big).
$$

%
%


%
%
%
%
%

\subsection{Estimation of $\alpha$}\label{ML}

If there is not enough information on $\alpha$ to fix a priori its value,
one option is to introduce a grid of values and repeat the MCMC scheme for any candidate.
Each value of $\alpha$ is a different model that, according to the Bayesian paradigm,
can be interpreted as a (discrete) random variable. One can then select
the structure having the largest posterior probability exploiting the 
MCMC outcomes \cite{Raftery}. 
The approach we propose relies on this strategy
and has connection with the concept of Bayesian evidence \cite{MacKayNC92}.
All the candidate values of $\alpha$ are seen as equiprobable and the 
(maximized) marginal likelihood is used
to approximate the posterior probabilites.\\
In dynamic networks, for a given $\alpha$, our optimized marginal likelihood is the total probability 
of $Y, \{\theta_k,\lambda_k\}_{k=1}^p,\sigma^2$ and $\tau$ 
where the modules are integrated-out and the variances are set
to the maximizers found among the MCMC samples.
Since $Y$ and $\{\theta_k\}_{k=1}^p$ conditional on the other variables
are jointly Gaussian and linearly related, 
${\bf p}(Y \big | \sigma^2,\{\lambda_k\}_{k=1}^p,\tau)$ is easily defined by
$$
Y \big | \big(\sigma^2,\{\lambda_k\}_{k=1}^p,\tau \big) \sim \mathcal{N}\big(0,\Sigma_Y\big)
$$    
where
$$
\Sigma_Y = G  \Sigma G^\top 
+ \sigma^2 I_{mp},
$$
with
$$
G=[G_1 \ldots G_p], \ \
\Sigma=\text{blkdiag}(\tau^2 \lambda^2_1K,\ldots,\tau^2 \lambda^2_p K).
$$

%
The marginal likelihood is then fully defined by
\begin{equation}\label{eqML}
{\bf p}(Y \big | \sigma^2,\{\lambda_k\}_{k=1}^p,\tau)  \sigma^{-2} {\bf p}(\tau) \prod_{k=1}^p {\bf p}(\lambda_k)
\end{equation}
with the priors on $\tau$ and $\lambda_k$ given by the half-Cauchy distributions $\mathcal{C}^+(0,1)$.
Since, as said, the variances are set to the optimizers, the above expression becomes only function of
the unknown  $\alpha$.\\ 
If the size of $Y$ is large, marginal likelihood evaluation can be expensive but this 
operation can be performed only 
on a subsample of the chain, e.g. once every 50-100 iterations.


\section{Numerical experiments}\label{Section5}

The non null impulse responses used during the numerical experiments
are randomly generated rational transfer functions
of order 10.
The  roots of the numerator and denominator are selected as follows.  
With the same probability a real or a couple of complex conjugate poles is added to the denominator until its order reaches  10.  Real poles are drawn from a uniform distribution on $[-0.95,0.95]$ 
while the complex conjugate pairs are obtained by using
uniform distributions on $[0,0.95]$ 
and $[0,\pi]$ for the absolute values and phases, respectively. 
Finally, each impulse response so generated is normalized
so that its Euclidean norm becomes a uniform random variable over
$[0.2,1]$.\\
In what follows, given an impulse response estimate $\hat{\theta}$ of the true 
and non null $\theta$, 
the estimation performance will be measured by the fit given by
\begin{equation}\label{eqfit}
100\%\left(1-\frac{\|\theta-\hat{\theta}\|}{\|\theta\|}\right).
\end{equation}
We will also compute the norm of the estimates of the 
null impulse responses: small values will indicate that 
the estimator has detected that such modules 
are not significant.\\
Finally, the starting point for the MCMC scheme is in any case
given by impulse responses containing coefficients all equal to $10^{-4}$.

\subsection{Sparse network estimation using white noise or low-pass inputs}

\begin{figure}
\center {\includegraphics[scale=0.4]{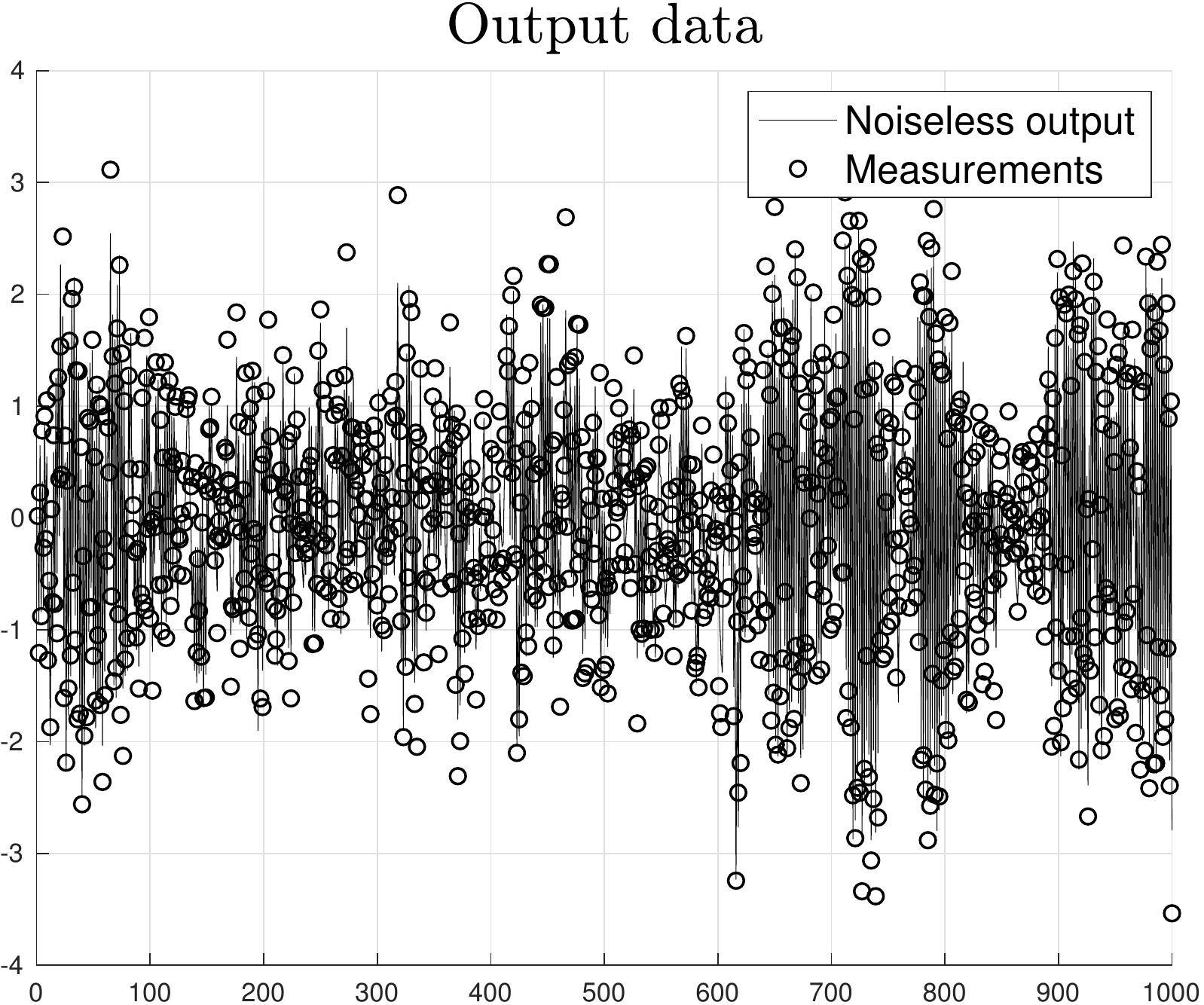}}
\caption{{\bf Use of white noises as inputs:} noiseless output and measurements (SNR is 10).}
\label{Fig3}
\end{figure}

We consider a simulated network containing 50 modules.
They are all null except the first 3.
In the first experiment the inputs $u(t)$ are independent realizations from white Gaussian noises of unit variance.
The signal to noise ratio (SNR) equal to 10 and,
after getting rid of initial conditions effects, $n=1000$ outputs forming the vector $Y$
are collected, see Fig. \ref{Fig3}. 
The modules are described by FIR models of 
dimension $m=200$. Hence, 10000 impulse response coefficients 
have to be estimated from the 1000 output data.\\
The variance decay rate $\alpha$ of the stable spline kernel is set to $0.9$
which roughly corresponds to believe that an upper bound of the dominant 
network poles is around $0.95$.
Using the MCMC scheme described in the previous section
200000 posterior samples are generated.
Fig. 4 plots the realizations of the global (left panel) and
of the 50 local shrinkage parameters (right).
Estimate of $\tau^2$ is really small, indicating that the level of 
network sparsity is large. The right panel shows that 
three local variances are counteracting the attraction 
to zero, corresponding to the three non null impulse responses.
This can be seen in Fig.  \ref{Fig5} that reports the minimum variance estimates of the 50 modules.
The first three are quite close to truth while all the other ones 
are virtually set to zero. 
In particular, the three panels in Fig.  \ref{Fig6} provide a
much better visualization of the estimates of the $3$ non null impulse responses:
the fits are $88.1\%,88.2\%$ and $80.4\%$. 
Finally, the right bottom panel of Fig. \ref{Fig6} 
contains the boxplot of the estimated norms of the $47$ null impulse responses.
Their values are all really small, indicating correctly that the corresponding links
are irrelevant for network dynamics.

\begin{figure*}
\center {\includegraphics[scale=0.4]{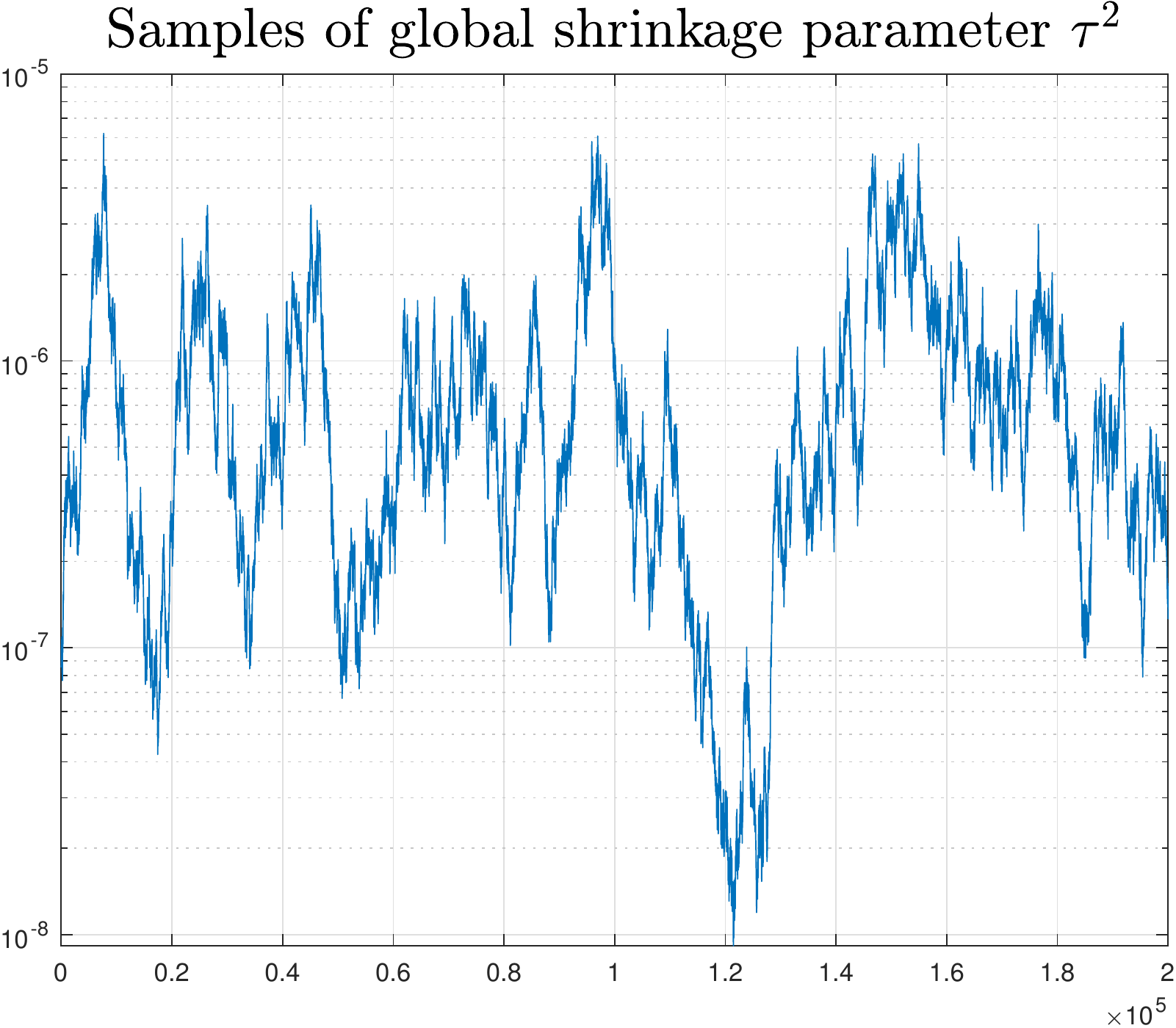}} \  {\includegraphics[scale=0.4]{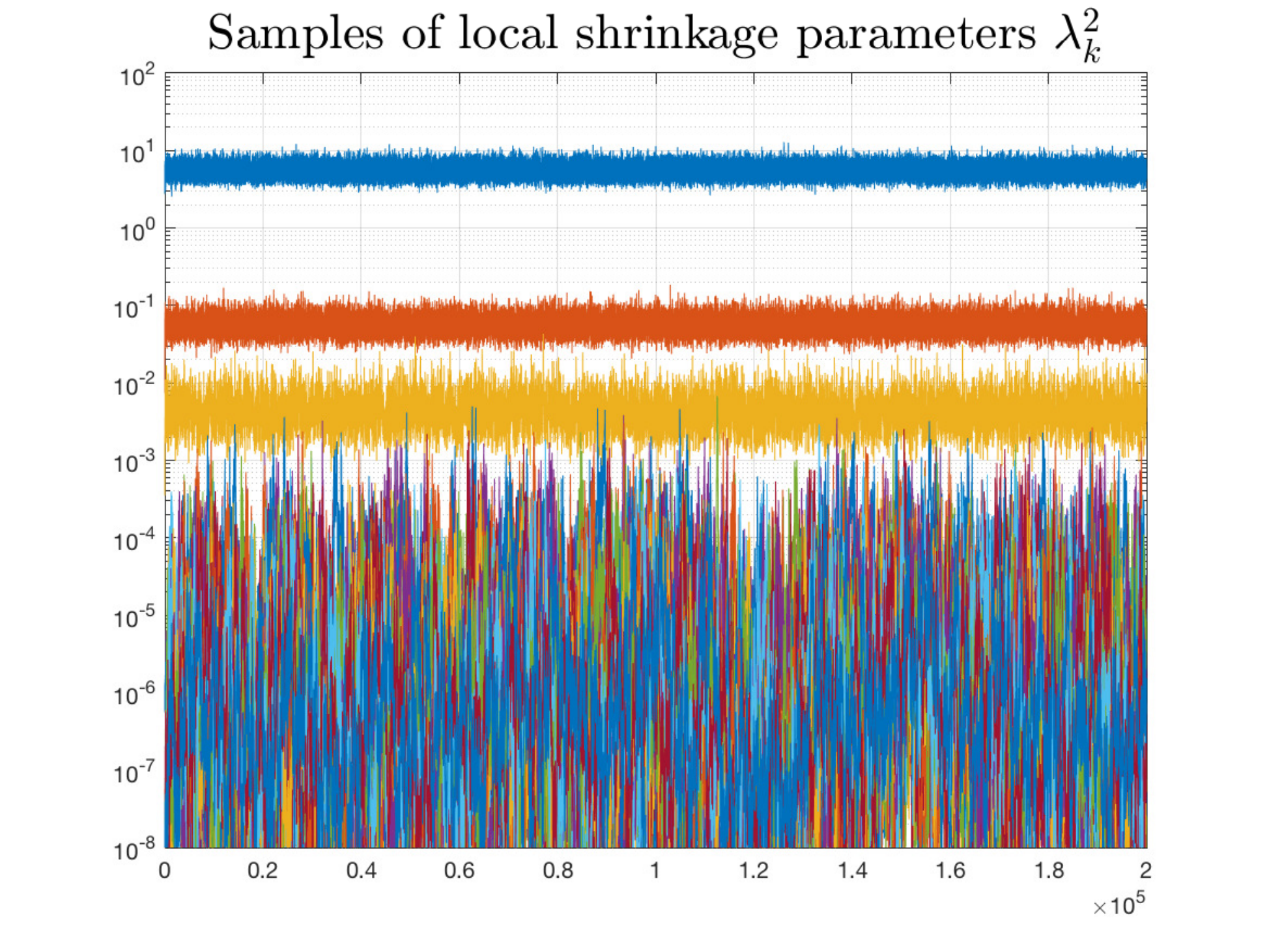}} \\
\caption{{\bf Use of white noises as inputs:} samples from the posterior of the global shrinkage parameter $\tau^2$ (left) 
and of the $50$ local shrinkage parameters $\lambda_k^2$ (right panel, where the three trajectories containing the largest variance values are 
associated to the three non null impulse responses).}
\label{Fig4}
\end{figure*}


\begin{figure}
\center {\includegraphics[scale=0.4]{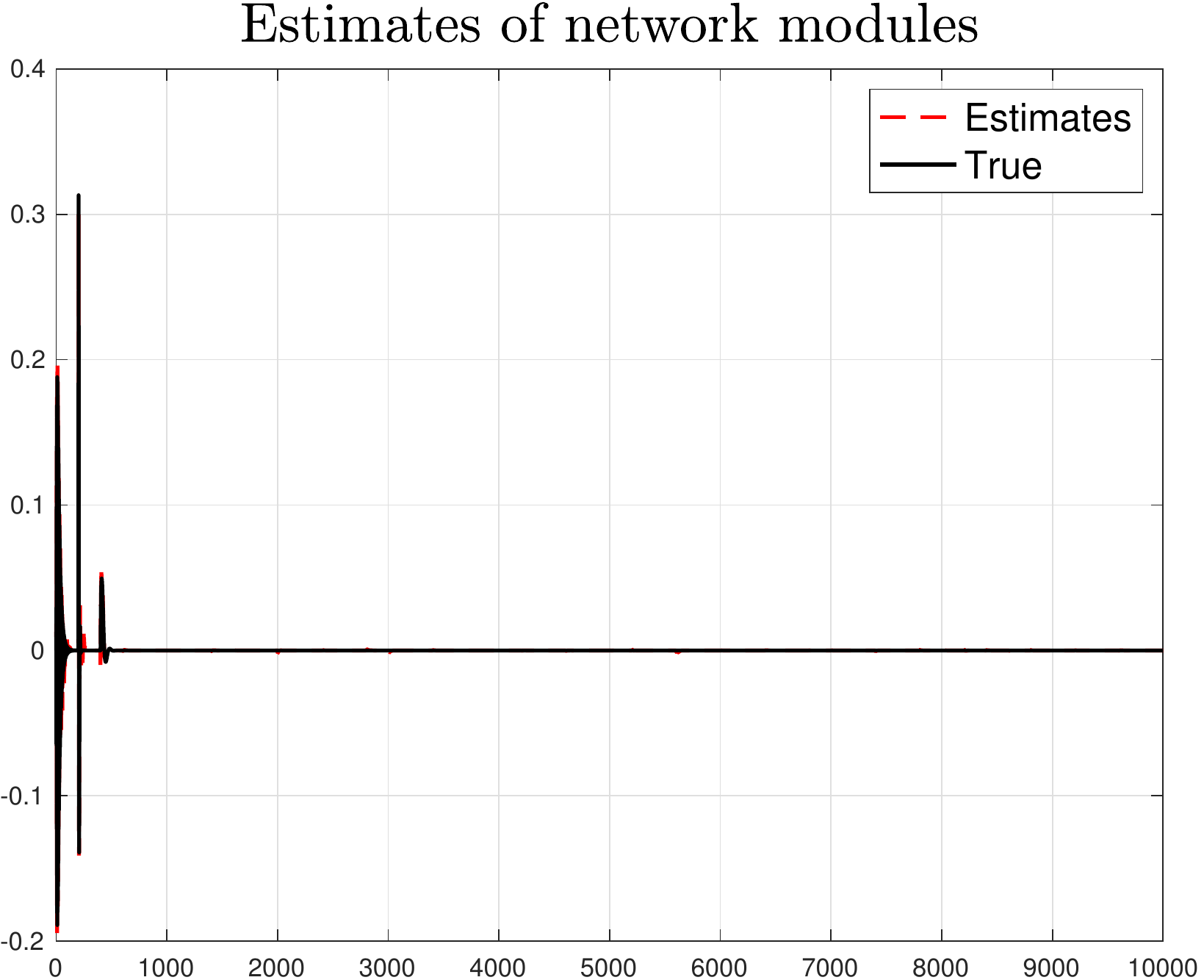}}
\caption{{\bf Use of white noises as inputs:} true (solid) and estimated (dashed) modules.
Only the first three impulse responses are really different from zero.}
\label{Fig5}
\end{figure}

\begin{figure*}
\center {\includegraphics[scale=0.35]{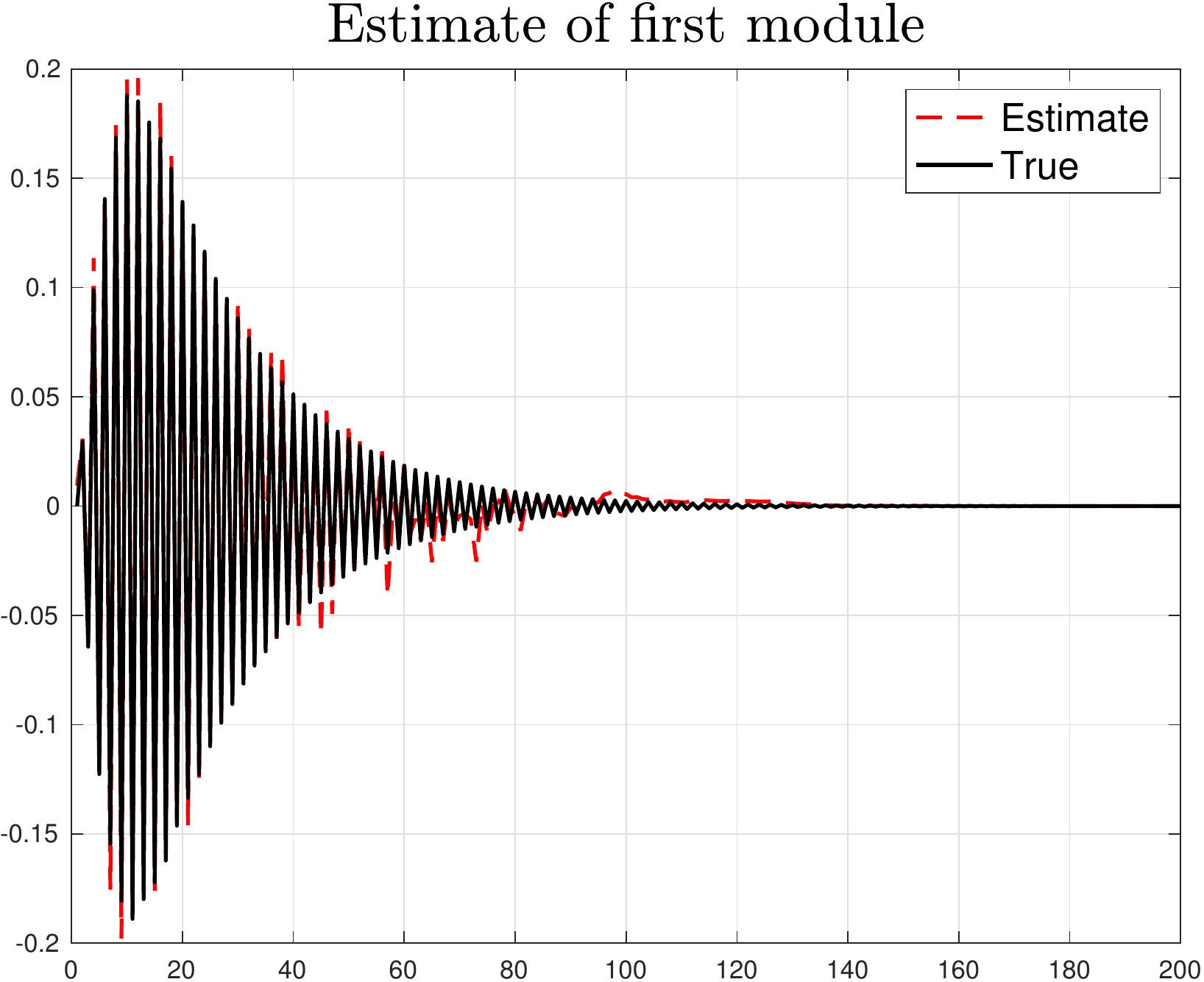}} \ {\includegraphics[scale=0.35]{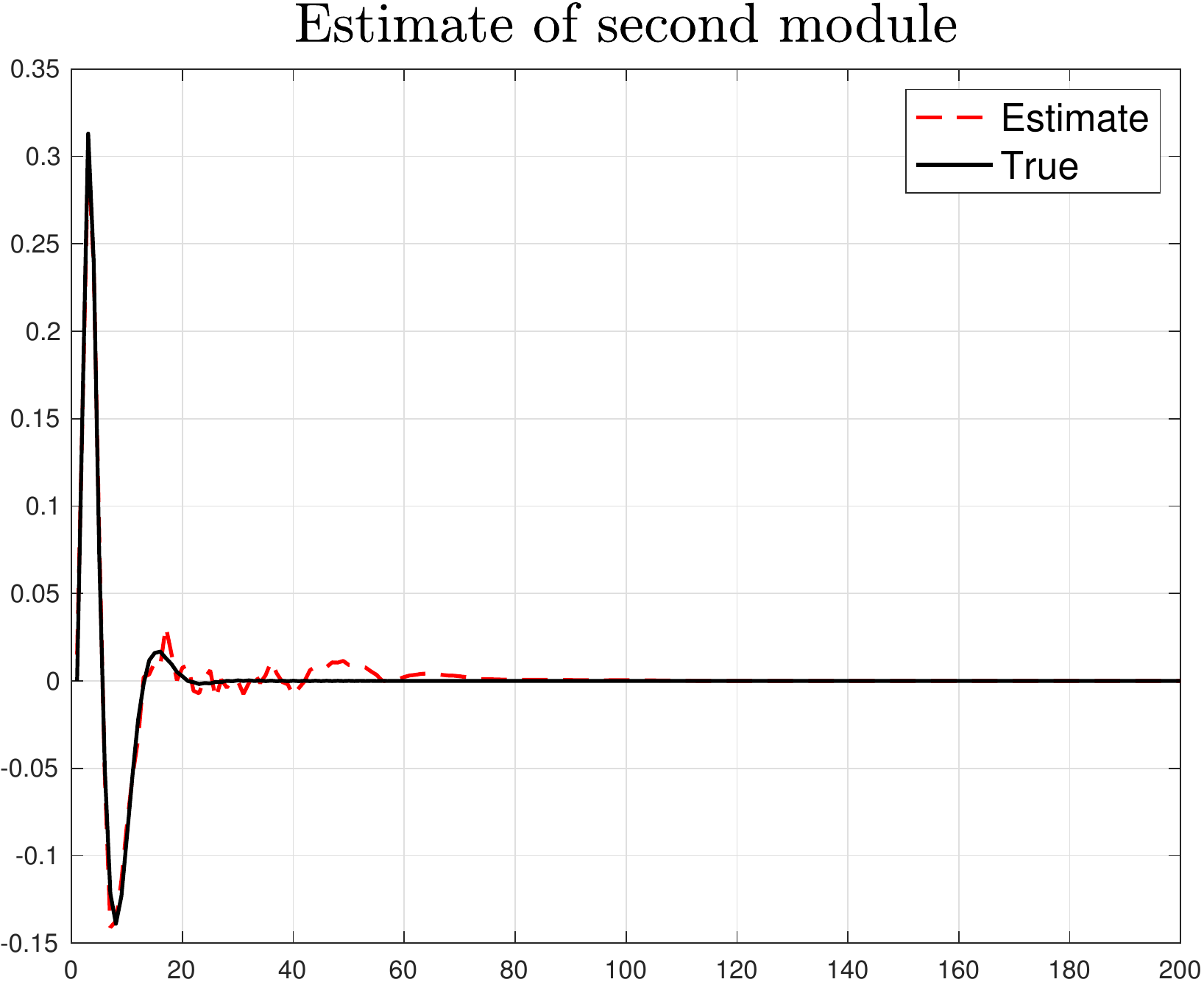}} \\ 
\center {\includegraphics[scale=0.35]{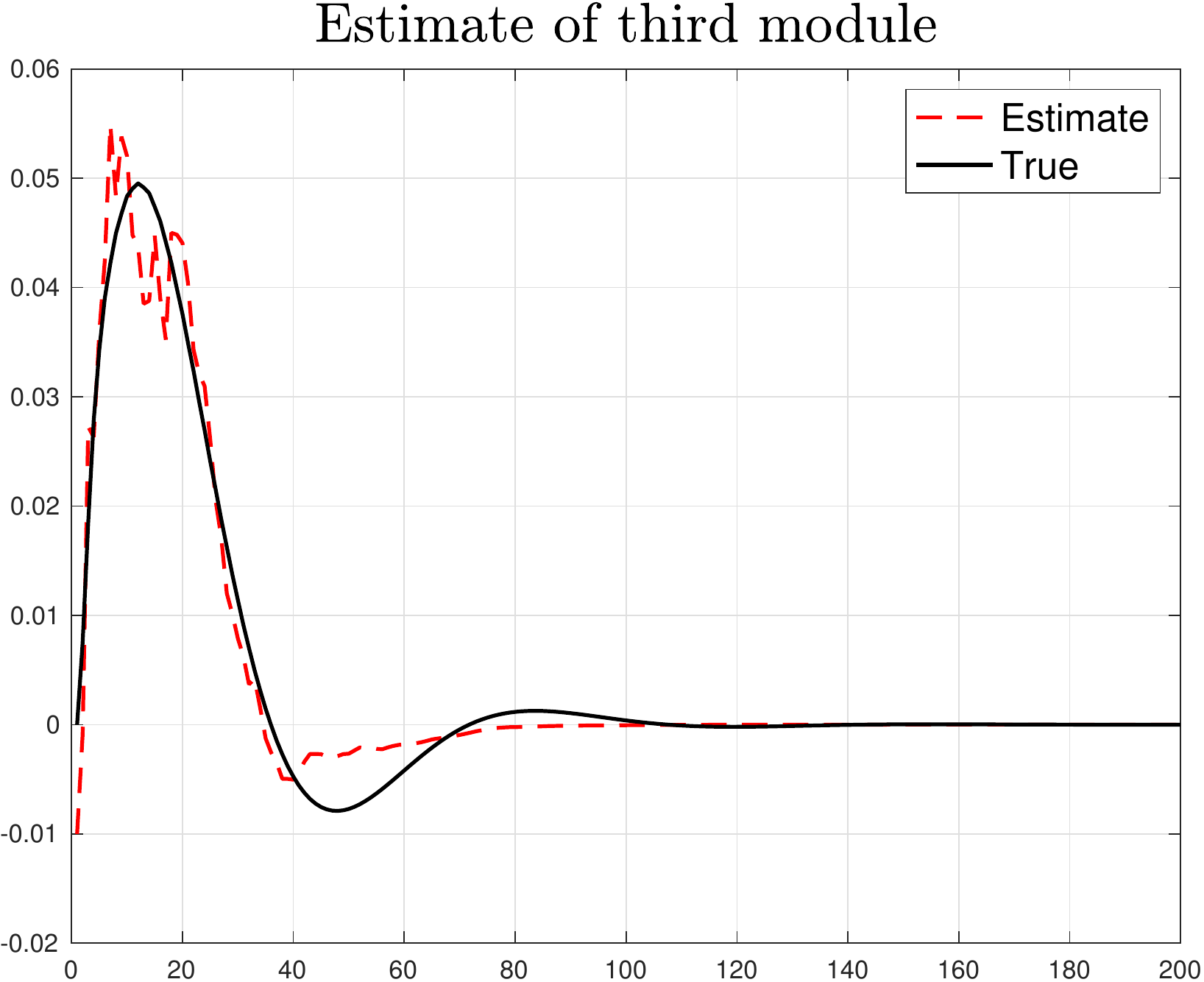}} \  {\includegraphics[scale=0.35]{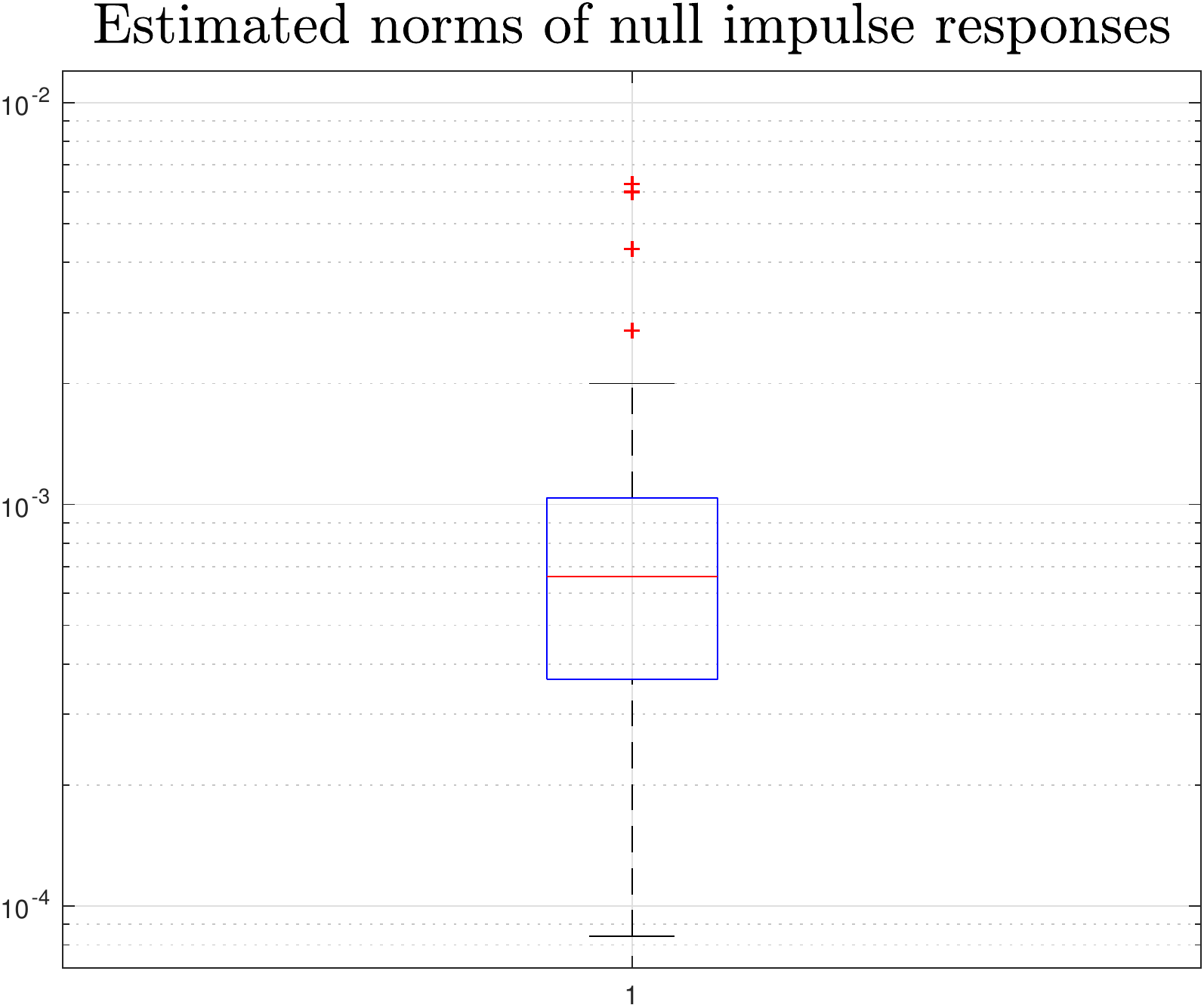}}
\caption{{\bf Use of white noises as inputs:} the first three panels show the estimates of the $3$ non null impulse responses.
The fits computed as in \eqref{eqfit} are $88.1\%,88.2\%$ and $80.4\%$
Finally, the bottom left panel displays the estimated norms of the remaining $47$ null impulse responses.}
\label{Fig6}
\end{figure*}


\begin{figure*}
\center {\includegraphics[scale=0.35]{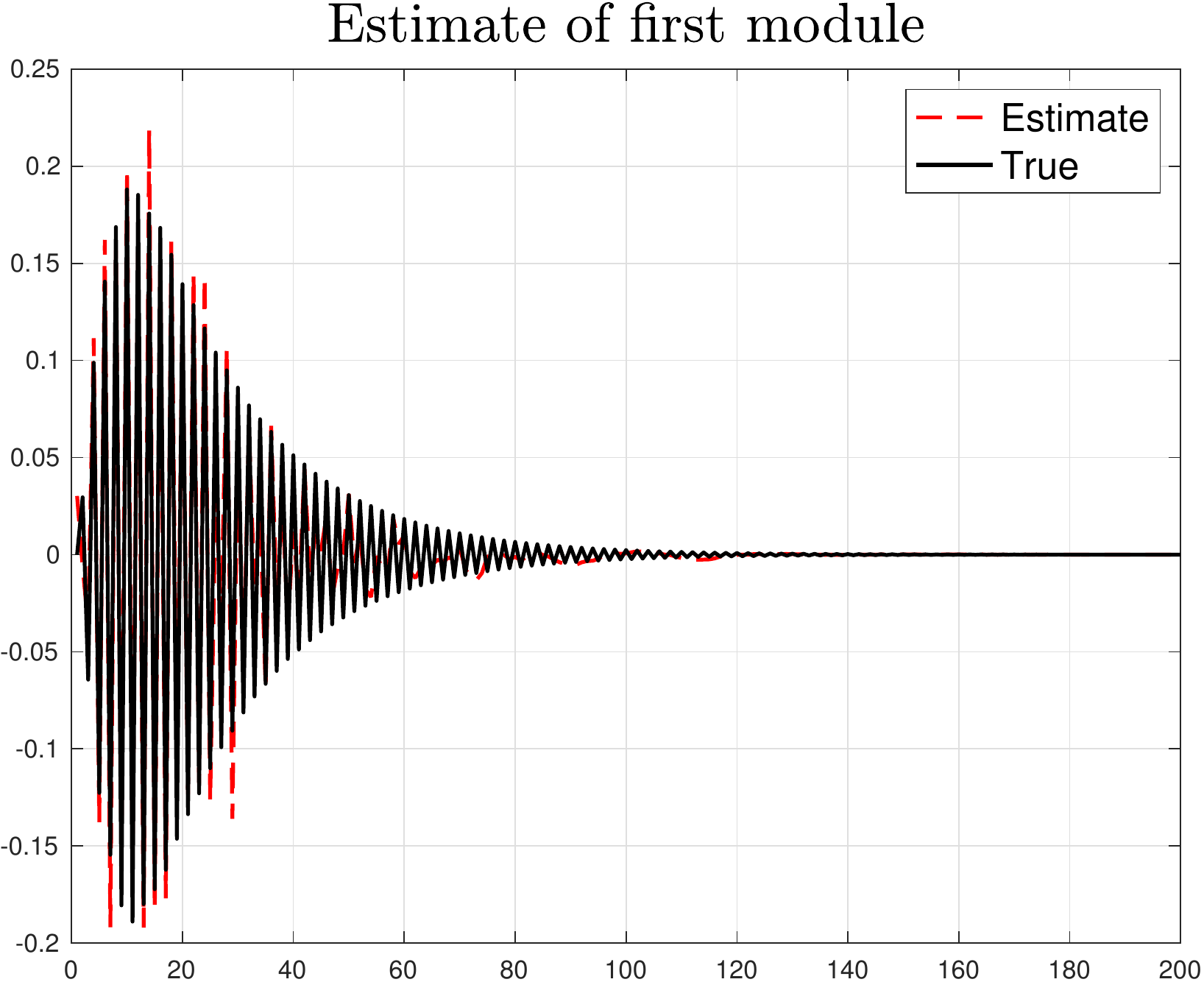}} \ {\includegraphics[scale=0.35]{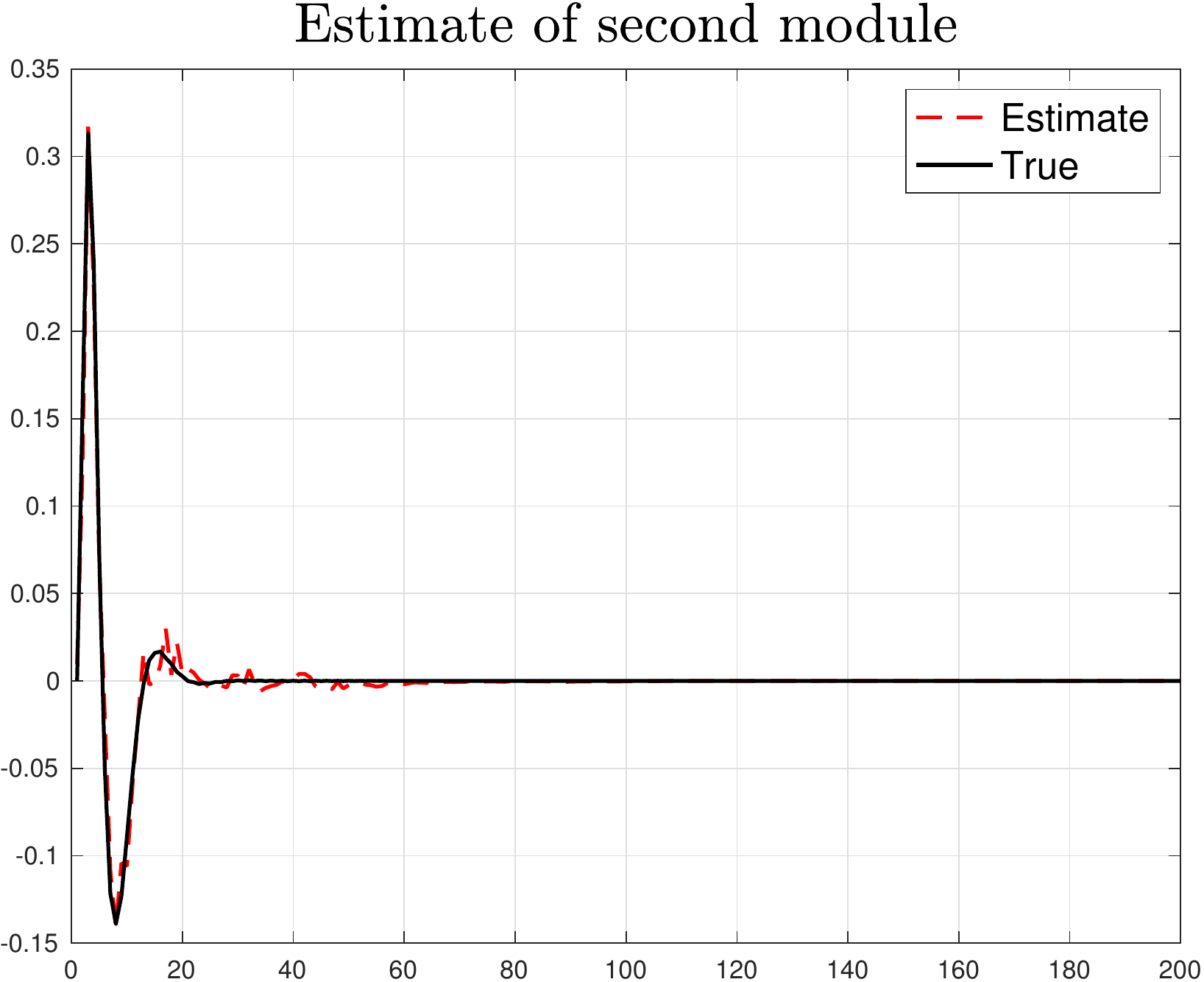}} \\ 
\center {\includegraphics[scale=0.35]{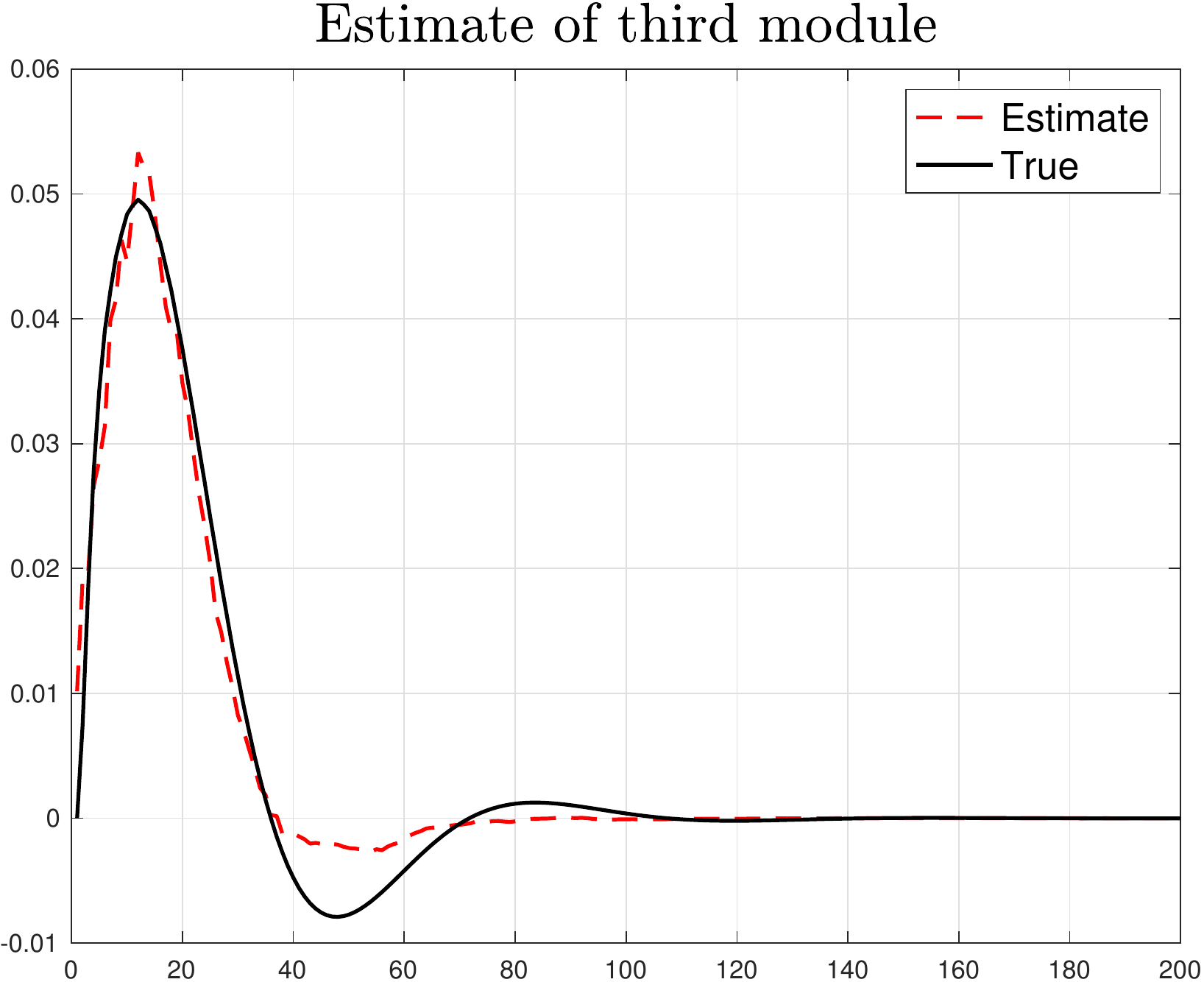}} \  {\includegraphics[scale=0.35]{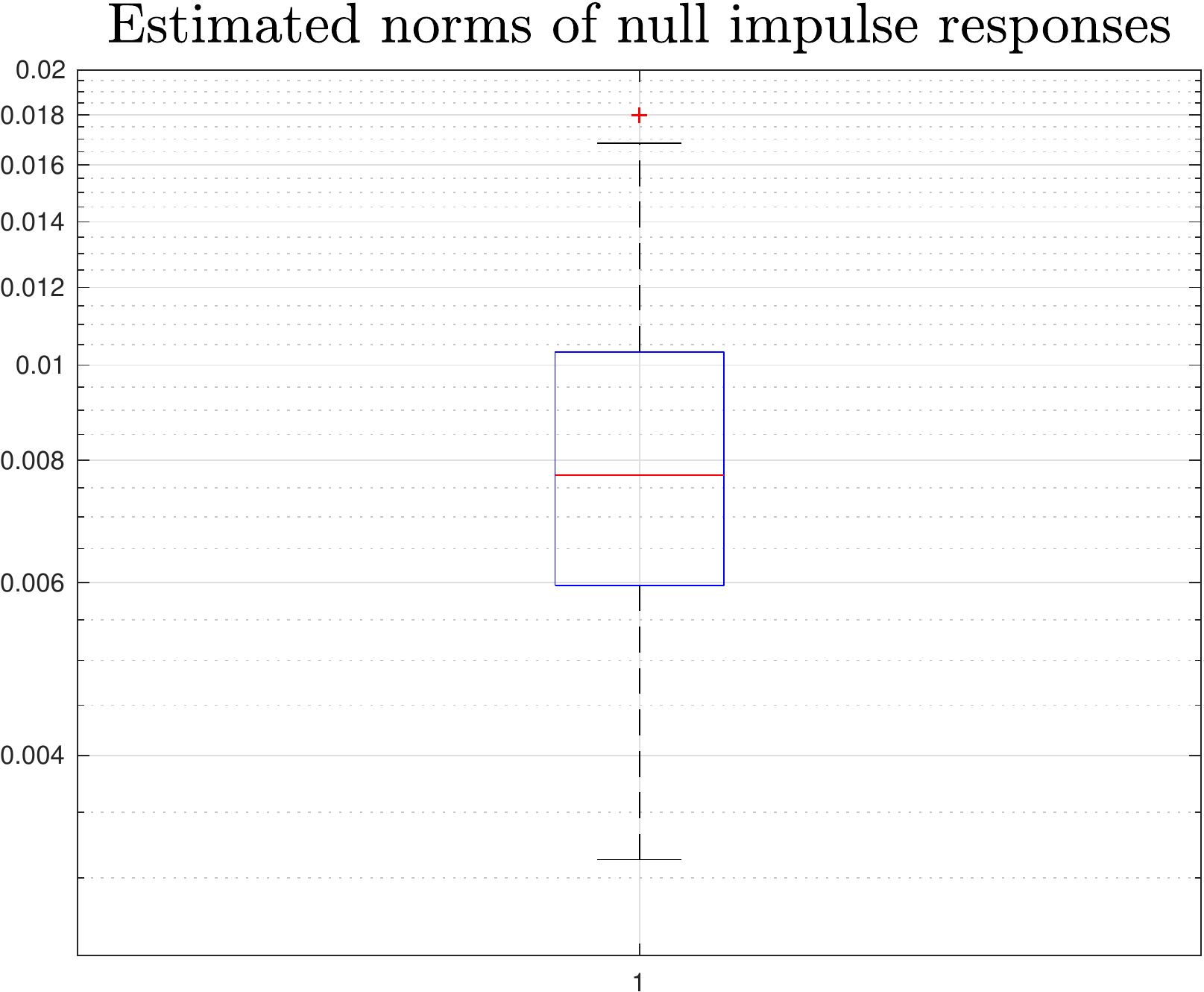}}
\caption{{\bf Use of low-pass inputs:} the first three panels show the estimates of the $3$ non null impulse responses.
The fits computed as in \eqref{eqfit} now turn out to be $70.9\%,85.7\%$ and $83.6\%$.  
Finally, the bottom left panel displays the estimated norms of the remaining $47$ null impulse responses.}
\label{Fig7}
\end{figure*}

We repeated the experiment using as inputs white Gaussian noise
filtered by a low-pass filter $1/(z-0.9)$, hence increasing the ill-conditioning
affecting the problem.
Fig.  \ref{Fig7} reports the same kind of results contained in Fig.  \ref{Fig6} .
The estimator still exhibits a high performance:
the fits of the three non null impulse responses are $70.9\%,85.7\%$ and $83.6\%$.
As expected, it is now more difficult to estimate the first impulse response since
its contents are more concentrated at high-frequencies where input power is small.
However, all the reconstructed profiles remain satisfactory.
Furthermore, the boxplot of the estimated norms of the other modules
still contains quite small values: also in this case the estimator
has detected that the remaining 47 modules are not so significant.\\

\begin{figure*}
\center {\includegraphics[scale=0.4]{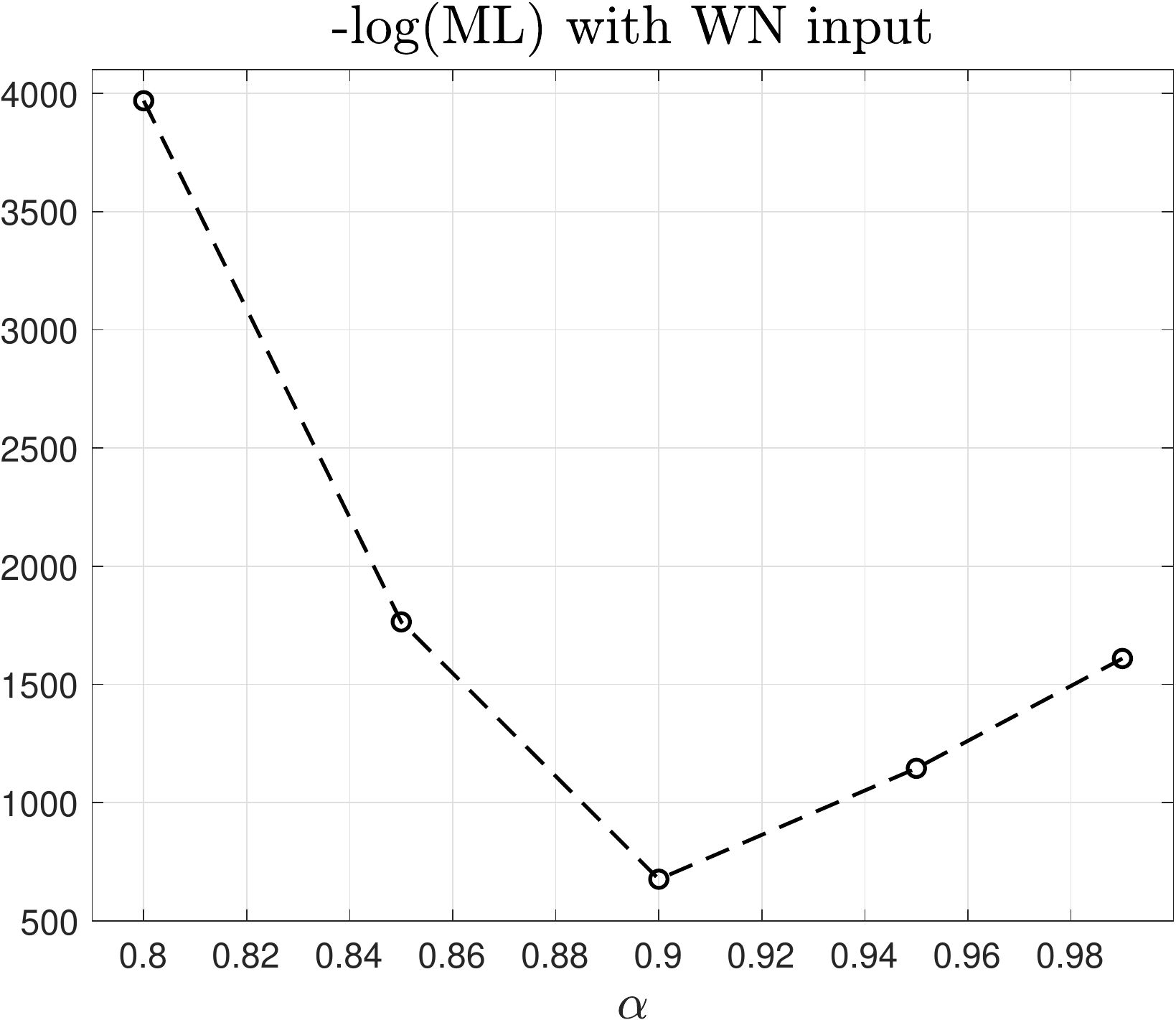}} \  {\includegraphics[scale=0.4]{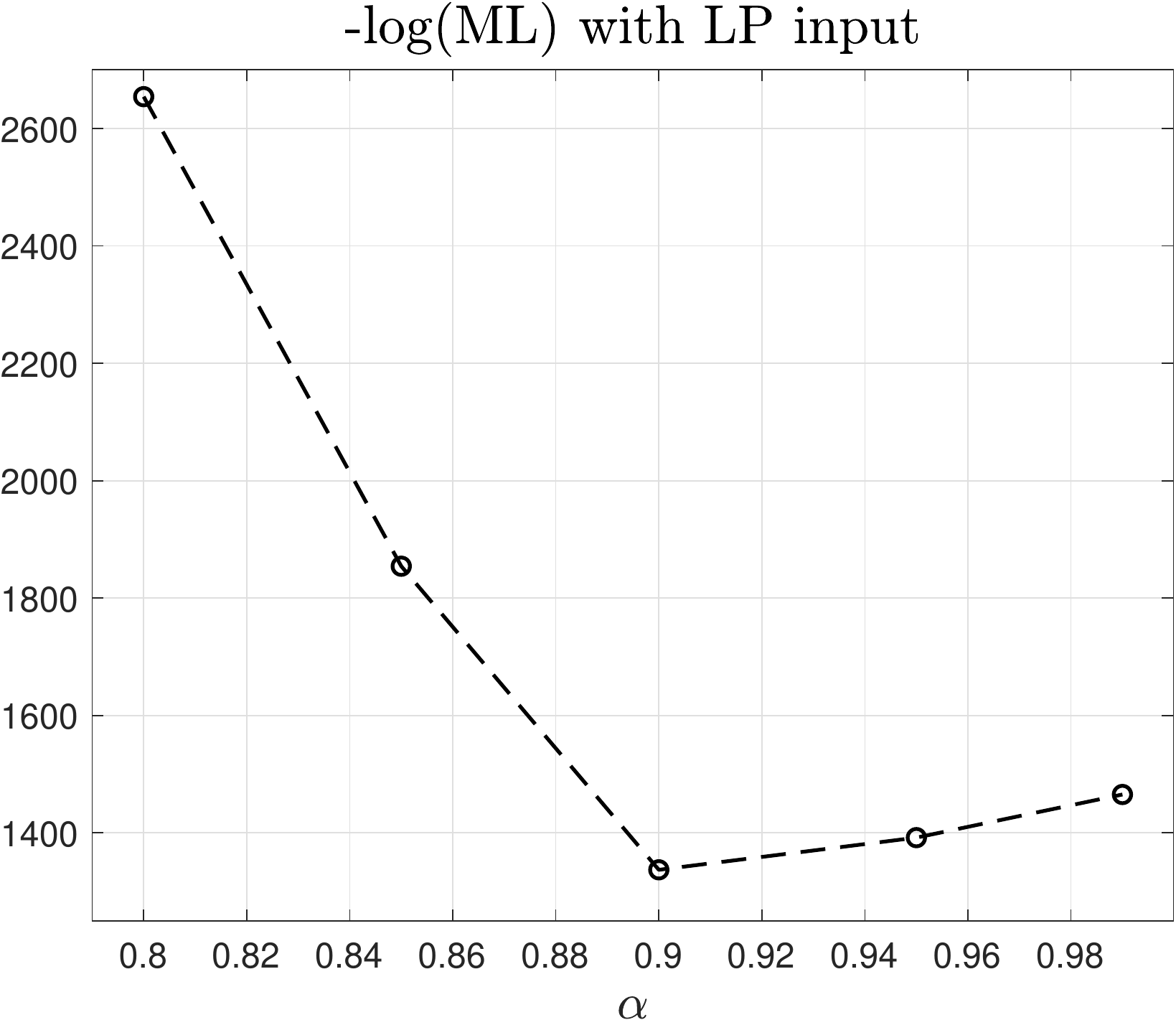}} \\
\caption{{\bf Selection of the impulse response variance decay rate $\alpha$:} optimized value
of the minus logarithm of the marginal likelihood \eqref{eqML} using white noise (left panel) or low pass (right panel) input.}
\label{FigML}
\end{figure*}

In both the two experiments we have also assessed which value of $\alpha$
would be estimated by adopting the marginal likelihood strategy described in 
section \ref{ML}. Using the grid $[0.8,0.85,\ldots,0.95,0.99$]
in both the cases we obtained the adopted value
$\alpha=0.9$. This is documented in Fig. \ref{FigML} which reports the optimized value
of the minus logarithm of \eqref{eqML} using white noise (left panel) or low pass (right panel) input.
So, in these experiments the marginal likelihood estimator does a nice job in 
detecting the upper bound on the absolute values of the network dominant poles.

\subsection{Monte Carlo study}

\begin{figure*}
\center {\includegraphics[scale=0.4]{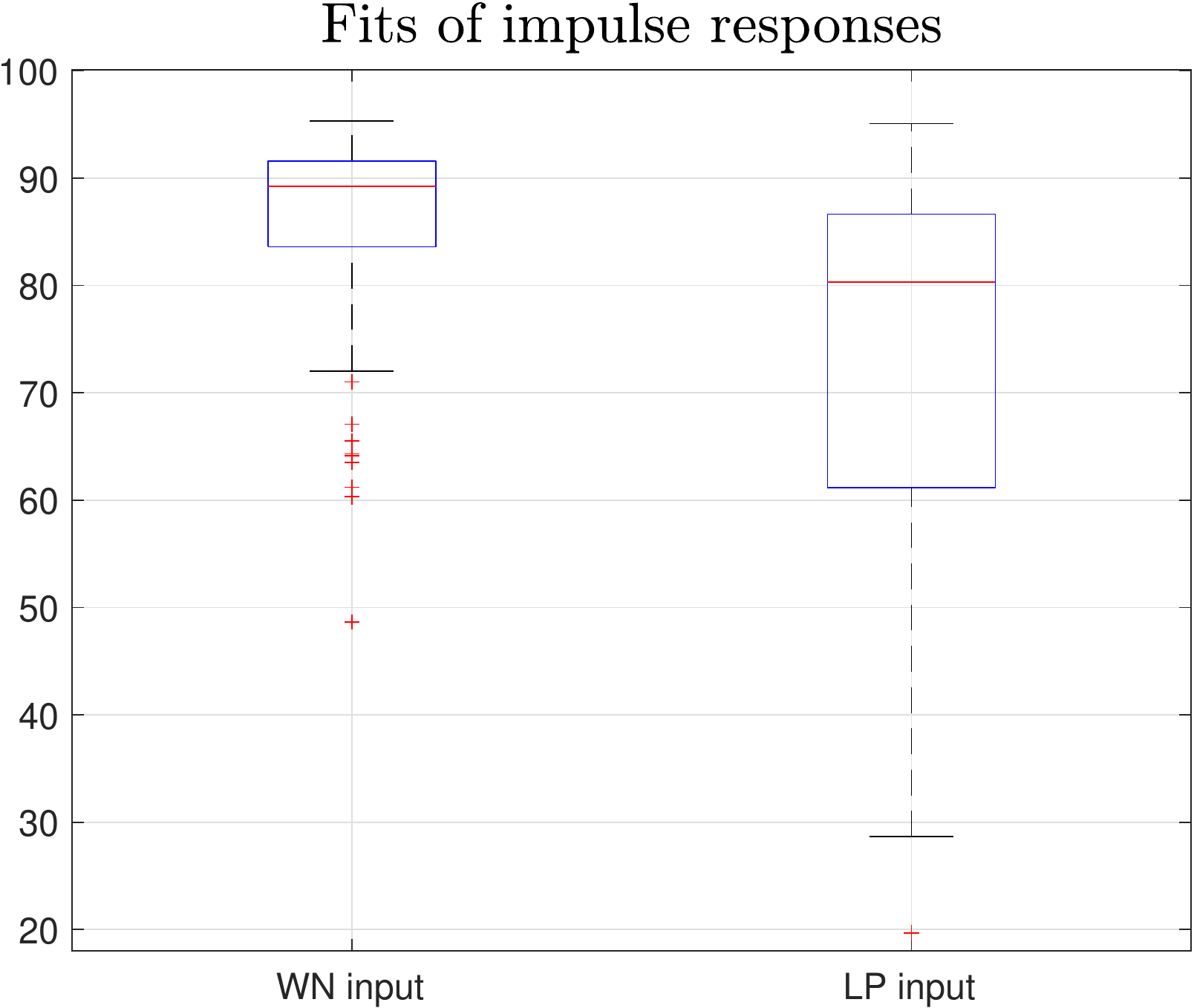}} \  {\includegraphics[scale=0.4]{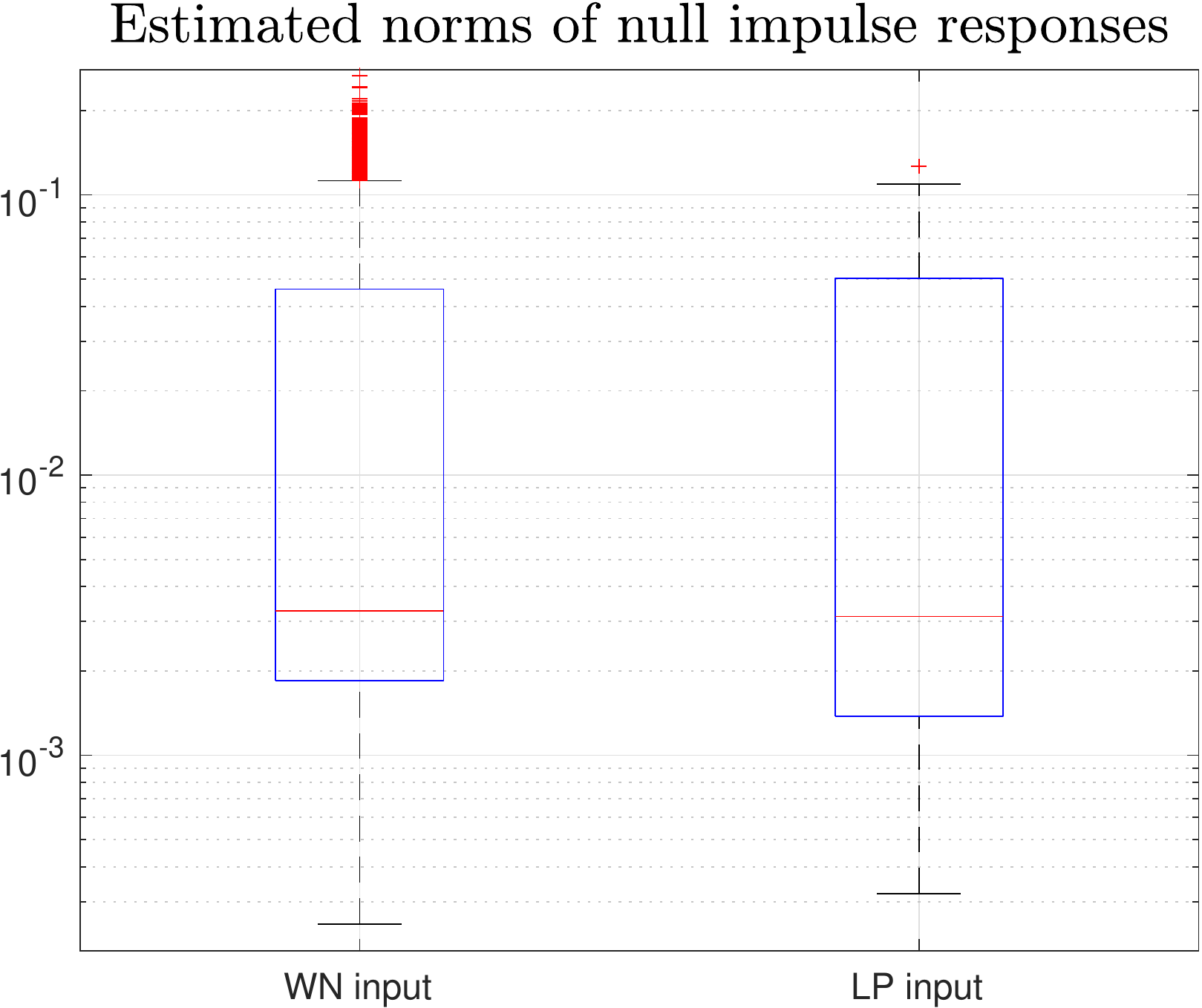}} \\
\caption{{\bf Monte Carlo study:} fits of impulse responses  (left panel) and estimated norms of impulse responses 
(right panel). In each panel  the left and right boxplot collect results obtained
using white noise or a low pass signal as input, respectively.}
\label{FigMC}
\end{figure*}

We now consider a randomized version
of the previous two case studies.
Simulated data are generated exactly as described in the previous section
except that at any run 
the number of non null impulse responses $q$
is a uniform random variable with values 
in $\{0,1,2,3\}$. The non null impulse responses are then assigned to
$q$ modules whose indexes are randomly selected among the 50 forming the network.
Note that the case $q=0$ corresponds to a network where all the connections are absent.
Two Monte Carlo studies of 100 runs
are so considered  with inputs given by white noises or 
low-pass signals. At any run we achieve
the fits of the non null impulse responses and 
the estimated norms of the null ones.
The boxplots reported in Fig. \ref{FigMC} report all of this information.
In the left panel one can see the fits of 139 impulse responses
in the case of white noise input. The estimator's performance is 
remarkable with the median close to $90\%$.
Still in the left panel one can also see the fits of 136 impulse responses in the
case of low pass input. The boxplot contains
a larger range of values extending also to small fits 
due to the presence of severe ill-conditioning.
However, the quality of the estimates is quite good,
with the median around $80\%$ and most of the fits
larger than $60\%$.\\
The right panel reports the boxplots with the estimated norms of 4861 (left)
and 4864 (right) impulse responses in the two cases.
All the contained values are quite small.
Recalling that the norms of the non null impulse responses
are uniform random variable between $[0.2,1]$, one can conclude that 
the stable spline horseshoe estimator
has a high capability to discriminate which networks links are operating.

%
%

\section{Conclusions}

We think that the combination of the horseshoe prior and the stable spline kernel
defines a powerful description of sparse dynamic networks.
In fact, many real interconnected systems can be well described by a two-groups model,
where many impulse responses are null while the other ones are exponentially stable.
Cast in a Bayesian framework, this model would lead to
a complex posterior distribution defined over a high-dimensional discrete space difficult to explore.
The numerical studies here reported show that the
stable spline horseshoe prior can well approximate the true posterior deriving from such two-groups mixture.
The reason is that the distributions of the scale factors of any local stable spline kernel are now defined by
half-Cauchy distributions. Hence, they inherit the features of the horseshoe prior,
inducing a distribution on the impulse response norm 
with flat tails and an infinitely tall spike
at zero.  Null modules can be aggressively shrunk towards zero while 
the significant ones can be retained and 
regularized by the stable spline prior.\\ 
The a posteriori density function defined by the new model 
can be reconstructed in sampled form by MCMC. 
By accounting for dynamic systems peculiarities, our Gibbs sampling scheme
turns out especially efficient, allowing to generate thousands of samples
from the modules posterior with low computational cost. 
Hence, one can obtain minimum variance estimates, 
as well as uncertainty bounds around them, permitting to detect 
which connections are really influencing network dynamics. 
In future studies we plan to extend this approach
to even more complex networks where modules can follow also nonlinear dynamics.


\begin{thebibliography}{10}

\bibitem{Aronszajn50}
N.~Aronszajn.
\newblock Theory of reproducing kernels.
\newblock {\em Trans. of the American Mathematical Society}, 68:337--404, 1950.

\bibitem{Bazanella2017}
A.S. Bazanella, M.~Gevers, J.M. Hendrickx, and A.~Parraga.
\newblock Identifiability of dynamical networks: Which nodes need be measured?
\newblock In {\em 2017 IEEE 56th Annual Conference on Decision and Control
  (CDC)}, pages 5870--5875, 2017.

\bibitem{AbsSum2020}
M.~Bisiacco and G.~Pillonetto.
\newblock Kernel absolute summability is sufficient but not necessary for rkhs
  stability.
\newblock {\em SIAM journal on control and optimization}, 2020.

\bibitem{MathFoundStable2020}
M.~Bisiacco and G.~Pillonetto.
\newblock On the mathematical foundations of stable rkhss.
\newblock {\em Automatica}, 2020.

\bibitem{Carvalho2009}
C.~Carvalho, N.~Polson, and J.~Scott.
\newblock Handling sparsity via the horseshoe.
\newblock In David van Dyk and Max Welling, editors, {\em Proceedings of the
  Twelfth International Conference on Artificial Intelligence and Statistics},
  volume~5 of {\em Proceedings of Machine Learning Research}, pages 73--80,
  Hilton Clearwater Beach Resort, Clearwater Beach, Florida USA, 16--18 Apr
  2009.

\bibitem{Carvalho2010}
C.~Carvalho, N.~Polson, and J.~Scott.
\newblock The horseshoe estimator for sparse signals.
\newblock {\em Biometrika}, 97(2):465--480, 2010.

\bibitem{CALCP14}
T.~Chen, M.~S. Andersen, L.~Ljung, A.~Chiuso, and G.~Pillonetto.
\newblock System identification via sparse multiple kernel-based regularization
  using sequential convex optimization techniques.
\newblock {\em IEEE Transactions on Automatic Control}, 59(11):2933--2945,
  2014.

\bibitem{CACCLP16}
T.~Chen, T.~Ardeshiri, F.P. Carli, A.~Chiuso, L.~Ljung, and G.~Pillonetto.
\newblock Maximum entropy properties of discrete-time first-order stable spline
  kernel.
\newblock {\em Automatica}, 66:34 -- 38, 2016.

\bibitem{ChenOL12}
T.~Chen, H.~Ohlsson, and L.~Ljung.
\newblock On the estimation of transfer functions, regularizations and
  {G}aussian processes - revisited.
\newblock {\em Automatica}, 48(8):1525--1535, 2012.

\bibitem{Chen19}
Tianshi Chen.
\newblock Continuous-time {DC} kernel --- a stable generalized first-order
  spline kernel.
\newblock {\em IEEE Transactions on Automatic Control}, 63:4442--4447, 2019.

\bibitem{Chiuso2012}
A.~Chiuso and G.~Pillonetto.
\newblock A {B}ayesian approach to sparse dynamic network identification.
\newblock {\em Automatica}, 48(8):1553--1565, 2012.

\bibitem{Dankers2016}
A.G. Dankers, P.M.J. {Van den Hof}, P.S.C. Heuberger, and X.~Bombois.
\newblock Identification of dynamic models in complex networks with prediction
  error methods: Predictor input selection.
\newblock {\em IEEE Transactions on Automatic Control}, 61(4):937--952, 2016.

\bibitem{Darwish2018}
M.A.H. Darwish, G.~Pillonetto, and R.~Toth.
\newblock The quest for the right kernel in {B}ayesian impulse response
  identification: The use of {OBFs}.
\newblock {\em Automatica}, 87:318 -- 329, 2018.

\bibitem{Donoho1992}
D.L. Donoho, I.M. Johnstone, J.C. Hoch, and A.S. Stern.
\newblock Maximum entropy and the nearly black object.
\newblock {\em J. Roy. Statist. Soc. Ser. B}, 54(1):41--81, 1992.

\bibitem{LARS2004}
B.~Efron, T.~Hastie, L.~Johnstone, and R.~Tibshirani.
\newblock Least angle regression.
\newblock {\em Annals of Statistics}, 32:407--499, 2004.

\bibitem{Everitt2018}
N.~Everitt, M.~Galrinho, and H.~Hjalmarsson.
\newblock Open-loop asymptotically efficient model reduction with the
  steiglitz-mcbride method.
\newblock {\em Automatica}, 89:221--234, 2018.

\bibitem{Fonken2020}
S.J.M. Fonken, M.~Ferizbegovic, and H.~Hjalmarsson.
\newblock Consistent identification of dynamic networks subject to white noise
  using weighted null-space fitting.
\newblock In {\em Proc. 21st IFAC World Congress, Berlin, Germany}, 2020.

\bibitem{Gilks}
W.R. Gilks, S.~Richardson, and D.J. Spiegelhalter.
\newblock {\em Markov chain Monte Carlo in Practice}.
\newblock London: Chapman and Hall, 1996.

\bibitem{Goncalves2008}
J.~Goncalves and S.~Warnick.
\newblock Necessary and sufficient conditions for dynamical structure
  reconstruction of lti networks.
\newblock {\em IEEE Transactions on Automatic Control}, 53(7):1670--1674, 2008.

\bibitem{Hagmann2008}
P.~Hagmann, L.~Cammoun, X.~Gigandet, R.~Meuli, C.J. Honey, V.J. Wedeen, and
  O.~Sporns.
\newblock Mapping the structural core of human cerebral cortex.
\newblock {\em {PLOS} Biology}, 6(7):1--15, 2008.

\bibitem{Hastie01}
T.~J. Hastie, R.~J. Tibshirani, and J.~Friedman.
\newblock {\em The Elements of Statistical Learning. Data Mining, Inference and
  Prediction}.
\newblock Springer, Canada, 2001.

\bibitem{Hayden2016}
D.~Hayden, Y.~{Hwan Chang}, J.~Goncalves, and C.J. Tomlin.
\newblock Sparse network identifiability via compressed sensing.
\newblock {\em Automatica}, 68:9--17, 2016.

\bibitem{Hendrickx2019}
J.M. Hendrickx, M.~Gevers, and A.S. Bazanella.
\newblock Identifiability of dynamical networks with partial node measurements.
\newblock {\em IEEE Transactions on Automatic Control}, 64(6):2240--2253, 2019.

\bibitem{Hickman2017}
R.~Hickman, M.C.~Van Verk, A.J.H. {Van Dijken}, M.P. Mendes, I.~A. Vroegop-Vos,
  L.~Caarls, M.~Steenbergen, I.~{Van der Nagel}, G.J. Wesselink, A.~Jironkin,
  A.~Talbot, J.~Rhodes, M.~{De Vries}, R.C. Schuurink, K.~Denby, C.M.J.
  Pieterse, and S.C.M. {Van Wees}.
\newblock Architecture and dynamics of the jasmonic acid gene regulatory
  network.
\newblock {\em The Plant Cell}, 29(9):2086--2105, 2017.

\bibitem{Jin2020}
J.~Jin, Y.~Yuan, and J.~Goncalves.
\newblock High precision variational bayesian inference of sparse linear
  networks.
\newblock {\em Automatica}, 118:109017, 2020.

\bibitem{Johndrow2020}
J.~Johndrow, P.~Orenstein, and A.~Bhattacharya.
\newblock Scalable approximate mcmc algorithms for the horseshoe prior.
\newblock {\em Journal of Machine Learning Research}, 21(73):1--61, 2020.

\bibitem{Raftery}
R.E. Kass and A.E. Raftery.
\newblock Bayes factors.
\newblock {\em J. Amer. Statist. Assoc.}, 90:773--795, 1995.

\bibitem{Ljung:99}
L.~Ljung.
\newblock {\em System Identification - Theory for the User}.
\newblock Prentice-Hall, Upper Saddle River, N.J., 2nd edition, 1999.

\bibitem{MacKayNC92}
D.J.C. MacKay.
\newblock {B}ayesian interpolation.
\newblock {\em Neural Computation}, 4:415--447, 1992.

\bibitem{Makalic2016}
E.~Makalic and D.F. Schmidt.
\newblock A simple sampler for the horseshoe estimator.
\newblock {\em IEEE Signal Processing Letters}, 23(1):179--182, 2016.

\bibitem{Materassi2010}
D.~Materassi and G.~Innocenti.
\newblock Topological identification in networks of dynamical systems.
\newblock {\em IEEE Transactions on Automatic Control}, 55(8):1860--1871, 2010.

\bibitem{Materassi2020}
D.~Materassi and M.~V. Salapaka.
\newblock Signal selection for estimation and identification in networks of
  dynamic systems: a graphical model approach.
\newblock {\em IEEE Transactions on Automatic Control}, 65(10):4138--4153,
  2020.

\bibitem{Pagani2013}
G.A. Pagani and M.~Aiello.
\newblock The power grid as a complex network: A survey.
\newblock {\em Physica A: Statistical Mechanics and its Applications},
  392(11):2688--2700, 2013.

\bibitem{Park2008}
T.~Park and G.~Casella.
\newblock The {B}ayesian {L}asso.
\newblock {\em Journal of the American Statistical Association},
  103(482):681--686, 2008.

\bibitem{PillonettoHybrid}
G.~Pillonetto.
\newblock A new kernel-based approach to hybrid system identification.
\newblock {\em Automatica}, 70:21 -- 31, 2016.

\bibitem{Pillonetto2016}
G.~Pillonetto, T.~Chen, A.~Chiuso, G.~{De Nicolao}, and L.~Ljung.
\newblock Regularized linear system identification using atomic, nuclear and
  kernel-based norms: The role of the stability constraint.
\newblock {\em Automatica}, 69:137 -- 149, 2016.

\bibitem{SS2010}
G.~Pillonetto and G.~{De Nicolao}.
\newblock A new kernel-based approach for linear system identification.
\newblock {\em Automatica}, 46(1):81--93, 2010.

\bibitem{SurveyKBsysid}
G.~Pillonetto, F.~Dinuzzo, T.~Chen, G.~De Nicolao, and L.~Ljung.
\newblock Kernel methods in system identification, machine learning and
  function estimation: a survey.
\newblock {\em Automatica}, 50(3):657--682, 2014.

\bibitem{PillonettoTAC2021}
G.~Pillonetto and A.~Scampicchio.
\newblock Sample complexity and minimax properties of exponentially stable
  regularized estimators.
\newblock {\em IEEE Trans. Automat. Contr.}, 2021.

\bibitem{Polson2012}
N.G. Polson and J.G. Scott.
\newblock {On the Half-Cauchy Prior for a Global Scale Parameter}.
\newblock {\em Bayesian Analysis}, 7(4):887 -- 902, 2012.

\bibitem{Prando2020}
G.~Prando, M.~Zorzi, A.~Bertoldo, M.~Corbetta, M.~Zorzi, and A.~Chiuso.
\newblock Sparse dcm for whole-brain effective connectivity from resting-state
  fmri data.
\newblock {\em NeuroImage}, 208:116367, 2020.

\bibitem{Ramaswamy2021}
K.R. Ramaswamy, G.~Bottegal, and P.M.~J. {Van den Hof}.
\newblock Learning linear models in a dynamic network using regularized
  kernel-based methods.
\newblock {\em Automatica}, 129(109591), 2021.

\bibitem{Ramaswamy2021b}
K.R. Ramaswamy and P.M.~J. {Van den Hof}.
\newblock A local direct method for module identification in dynamic networks
  with correlated noise.
\newblock {\em IEEE Transactions on Automatic Control}, 2021.

\bibitem{Scholkopf01b}
B.~Sch\"{o}lkopf and A.~J. Smola.
\newblock {\em Learning with Kernels: Support Vector Machines, Regularization,
  Optimization, and Beyond}.
\newblock (Adaptive Computation and Machine Learning). MIT Press, 2001.

\bibitem{Soderstrom}
T.~S{\"o}derstr{\"o}m and P.~Stoica.
\newblock {\em System Identification}.
\newblock Prentice-Hall, 1989.

\bibitem{Lasso1996}
R.~Tibshirani.
\newblock Regression shrinkage and selection via the {LASSO}.
\newblock {\em Journal of the Royal Statistical Society, Series B.},
  58:267--288, 1996.

\bibitem{Tipping2001}
M.~Tipping.
\newblock Sparse bayesian learning and the relevance vector machine.
\newblock {\em Journal of Machine Learning Research}, 1:211--244, 2001.

\bibitem{VdH2013}
P.M.J. {Van den Hof}, A.G. Dankers, P.S.C. Heuberger, and X.~Bombois.
\newblock Identification of dynamic models in complex networks with prediction
  error methods: basic methods for consistent module estimates.
\newblock {\em Automatica}, 49(10):2994--3006, 2013.

\bibitem{Pas2014}
S.L. {van der Pas}, B.J.K. Kleijn, and A.W. {van der Vaart}.
\newblock The horseshoe estimator: Posterior concentration around nearly black
  vectors.
\newblock {\em Electronic Journal of Statistics}, 8(2):2585 -- 2618, 2014.

\bibitem{Weerts2018}
H.H.M. Weerts, P.M.~J. {Van den Hof}, and A.G. Dankers.
\newblock Prediction error identification of linear dynamic networks with
  rank-reduced noise.
\newblock {\em Automatica}, 98:256--268, 2018.

\bibitem{Weerts2018b}
H.M. Weerts, P.M.J. {Van den Hof}, and A.G. Dankers.
\newblock Identifiability of linear dynamic networks.
\newblock {\em Automatica}, 89:247--258, 2018.

\bibitem{Yue2021}
Z.~Yue, J.~Thunberg, W.~Pan, L.~Ljung, and J.~Goncalves.
\newblock Dynamic network reconstruction from heterogeneous datasets.
\newblock {\em Automatica}, 123:109339, 2021.

\end{thebibliography}
\end{document}